\documentclass[aps,prd,twocolumn,showpacs,preprintnumbers,amsmath,amssymb]{revtex4-1}
\usepackage{graphicx}
\usepackage{color}
\definecolor{purple}{rgb}{0.6,0,0.5}
\definecolor{verde}{rgb}{0,0.5,0}
\usepackage[colorlinks=true, pdfstartview=FitV, linkcolor=purple,
citecolor=blue,urlcolor=navyblue]{hyperref}
\usepackage{graphics,psfrag}
\usepackage{graphicx,psfrag}

\begin{document}

\title{CP violation: Dalitz interference, CPT and FSI}

\author{J.~H.~Alvarenga~Nogueira$^a$, I.~Bediaga$^b$, A.~B.~R.~Cavalcante$^b$, 
T.~Frederico$^a$, O.~Louren\c co$^c$}

\affiliation{$^a$Instituto Tecnol\'ogico de Aeron\'autica, DCTA, 12228-900, S\~ao Jos\'e 
dos Campos, SP, Brazil \\
$^b$Centro Brasileiro de Pesquisas F\'isicas, 22290-180, Rio de Janeiro, RJ, Brazil \\
$^c$Departamento de Ci\^encias da Natureza, Matem\'atica e Educa\c c\~ao, CCA, 
Universidade Federal de S\~ao Carlos, 13600-970, Araras, SP, Brazil}

\date{\today}

\begin{abstract}
Resonances and final state interactions (FSI) play a role in the formation of CP 
violation (CPV) constrained by CPT invariance. We provide a general formulation of CPV 
including resonances and FSI starting from the CPT constraint. Our discussion 
is elaborated within a simple $B$ decay model with the $\rho$ and $f_0(980)$ resonances 
plus a non resonant background including the $\pi\pi \to KK$ coupled amplitude. We 
consider few illustrative examples to show the interference patterns appearing in the CP 
asymmetry, namely, that from the $\rho$ resonance plus a non-resonant amplitude, 
and that from the interference of the $\rho$ and $f_0(980)$ resonances. We perform the 
fit of the CP asymmetry for the charmless three-body $B^\pm$ decay channel $B^{\pm}\to 
\pi^{\pm} \pi^+\pi^-$ and obtain as outcome the $B^{\pm}\to \pi^{\pm} K^+K^-$ for $\pi\pi$ 
channel asymmetry in the mass region below $1.6$~GeV in fair agreement with the new data 
LHCb data. Analogously, we also describe the CP asymmetry of the $B^\pm\to 
K^\pm\pi^+\pi^-$ decay, with that from the $B^\pm\to K^\pm K^+K^-$ channel obtained as 
output. As in the previous case, we also found agreement with LHCb experimental data.
\end{abstract}


\maketitle

\section{Introduction}

The exact CPT invariance implies the identity of the lifetime of a particle and its charge 
conjugate. Therefore, when the individual partial decay widths of CP conjugate channels  
are different, due to the CP violation (CPV), the other channels must have equal 
amount of CP violation, with opposite sign, such that the total width of the particle and 
antiparticle are equal. Final state interaction  must be the responsible to 
distribute the CP asymmetry among the different conjugate decay channels such as the sum 
of partial widths provides identical total width for the particle and its anti-particle.    

The common belief says that CPT invariance is not a practical constraint to be taken into 
account when computing CP violation in charmless $B$ decays. It is naively expected that 
CP violation is distributed among several different coupled final state channels  with 
two, three, four and more hadrons~\cite{ikaros-bigi}. Therefore, the CPT constraint should 
be realized, in principle, by summing up over many hadronic decay channels. However, since 
hadronic many-body rescattering effects are far from being understood, it is evident that 
this phenomenological hypothesis should be better understood before consider its important 
consequences for the description of the CP violation phenomena.

In an opposite direction to the common believe,  we shown in detail in a previous 
paper~\cite{BedPRD14}, that the coupling between several final states must be suppressed 
at least in some charmless three body $B$ decays. A quick observation to the event  
distribution of these decays shows that they are placed basically  at the boundary of the 
Dalitz plot, in a region dominated by light two-body resonances. This dominance can be 
directly associated with the non-perturbative regime at the $B$ mass 
energy~\cite{susanne}. In addition, by considering the observations made by early 
experiments~\cite{CERN-Munich,LASS} that the low mass $\pi\pi$ and $K\pi$ region is 
dominated  by the elastic regime, the number of coupled channel expected for charmless 
three-body $B$ decays should be very restricted. There is a significant exception to this 
general picture for the  two-body mass distribution: the S-wave $\pi\pi \to KK$ 
re-scattering process between $1$ to $1.6$~GeV~\cite{Cohen1980,ref7}, that was also 
discussed in detail in our previous paper. So, at least for this class of decays, the 
existence of few possibilities of coupled hadronic channels demands the explicit 
consideration of the important constraint from CPT invariance in the study of CP 
violation. Unfortunately little is known, either theoretically or experimentally, about 
the event  distribution for others coupled channels with four or more hadrons in the final 
state. However, by looking at the profusion of observed low mass vector-vector charmless 
$B$ decays, it seems that they are also dominated by low mass resonances~\cite{PDG}. 

The inclusive CP asymmetry recently released by the LHCb collaboration for four charged 
$B$ charmless three-body decays, gives a hint about the correlation among the asymmetries 
in the coupled decay channels~\cite{expnew}. Actually, the experimental results for 
strangeness $\Delta=1$ decays $A_{CP}(B^\pm\rightarrow K^{\pm}\pi^+\pi^-) = +0.025 \pm 
0.004 \pm 0.004 \pm 0.007$ and $A_{CP}(B^\pm\rightarrow K^\pm K^+K^-) = -0.036 \pm 0.004 
\pm 0.002 \pm 0.007$ multiplied by the observed number of events~\cite{expnew} and 
corrected by the branching fractions (see Ref.~\cite{PDG}), which represent approximately 
the efficiency not reported in the paper, show that the number of events involved in the 
CP asymmetry in one decay channel is approximately the same  in the other channel but 
with opposite sign. The same happens, within the experimental errors, to the coupled 
decay channels with strangeness $\Delta =0$. Taking into account the experimental values 
$A_{CP}(B^\pm\rightarrow\pi^\pm\pi^+\pi^-) = +0.058 \pm 0.008 \pm 0.009 \pm 0.007$ and 
$A_{CP}(B^\pm\rightarrow\pi^\pm K^+K^-) = -0.123 \pm 0.017 \pm 0.012 \pm 0.007$, the 
number of events observed in each channel, corrected in the same way as before by the 
branching fractions, show the same correlation as above, namely, a similar number of 
events involved in the CP asymmetry but with opposite signs. 

Beyond the inclusive CPV, three-body decays allow to observe this asymmetry in 
the Dalitz phase-space. In principle, we can expect CP violation in charmless three-body 
$B$ decays coming from three different types of interferences involving weak and strong 
phases: (i) direct CPV due to the interference of the tree and penguin amplitudes in the 
same intermediate state. This first kind of CPV is based in the BSS model~\cite{BSS}. (ii) 
CP asymmetry produced through the interference between two different final states with 
different weak phases coupled by the final state interaction. 
This second case is constrained by the CPT invariance and it was already well described in 
our previous work~\cite{BedPRD14}. (iii) CPV from the interference between 
two neighbor resonances in the Dalitz plot, which share the same-phase space region. 
This third CPV kind has been pointed out in previous theoretical 
\mbox{papers~\cite{Mirandizing1,Mirandizing2,Gronau2013,ref1,ref2,ref14}}.

The present study addresses the three types of dynamics related with CPV in the Dalitz 
plot discussed above and, more specifically, we look for signatures involving each one of 
these contributions in the charmless decay channels $B^{\pm}\to \pi^{\pm} \pi^+\pi^-$, 
$B^{\pm}\to \pi^{\pm} K^+K^-$, $B^{\pm}\to K^{\pm} \pi^+\pi^-$ and  $B^{\pm}\to K^{\pm} 
K^+K^-$. To perform this study, we use the recently published LHCb paper~\cite{expnew}, 
in particular, the $\pi^+\pi^-$ and $K^+K^-$ mass distribution from the difference 
between $B^{-}$ and $B^{+}$. These differences other than give directly the CP violation 
distribution in the phase-space, minimize substantially problems of acceptance and 
background not available in the LHCb paper. 

Our work develops a CP asymmetry formula including resonances and FSI 
based on the CPT constraint. The CP asymmetry is derived in lowest order in the strong 
interaction and decomposed in angular momentum. The explicit expression for the CP 
asymmetry is found for the decays channels $B^{\pm}\to 
\pi^{\pm} \pi^+\pi^-$ and $B^{\pm}\to K^{\pm} \pi^+\pi^-$. Our discussion is  
exemplified by considering, within the isobar model, the interference patterns for the CP 
asymmetry from the $\rho$ and $f_0(980)$ resonances plus a non resonant background 
including the  contribution of the $\pi\pi \to KK$ coupled channel amplitude. We use this 
formula to fit the recent $B^{\pm}\to \pi^{\pm} \pi^+\pi^-$ LHCb data~\cite{expnew} and 
obtain as outcome the $B^{\pm}\to \pi^{\pm} K^+K^-$ in qualitative agreement with data 
for the mass region below $1.6$~GeV. We apply the same procedure to the $B^{\pm}\to 
K^{\pm} \pi^+\pi^-$ and $B^{\pm}\to K^{\pm} K^+K^-$ decays. We also take into account the 
LHCb separation between positive and negative $\cos\theta$ values, that supplies new 
details which reveal the role of the vector meson resonance in building the CP asymmetry 
interference pattern. 

The work is organized as follows. A brief review of the CPT constraint for deriving the 
CP asymmetry  with FSI and our notation are provided in Sec.~\ref{secCPT}. The CP asymmetry formula
in leading order of the strong interaction and angular momentum decomposition is given in 
Sec.~\ref{sect-CPA}. The resonances are introduced in the CP asymmetry expression in 
Sec.~\ref{sec:rescpt}. The formula for the interfering resonant and non-resonant 
amplitudes in the asymmetry is developed in Sec.~\ref{sec:interf}. The analysis of the 
various CP asymmetry terms is given in Sec.~\ref{sec:analysis}. The results from the 
fitting procedure to the asymmetry in the $B^\pm\to\pi^\pm\pi^+\pi^-$ are shown in 
Sec.~\ref{sec:cpvfit1}, where we found that the CPT violating terms can be disregarded. We 
also show results for $B^\pm\to\pi^\pm K^+K^-$ decay with no new parameters. An analogous 
study is performed at Sec.~\ref{sec:cpvfit2} for $B^\pm\to K^\pm\pi^+\pi^-$ and $B^\pm\to 
K^\pm K^+K^-$ decays. The concluding remarks are provided in Sec.~\ref{sec:finalremarks}, 
where we also discuss a more general formula to the CP asymmetry, where the FSI terms 
interfere with the $\rho$ and $f_0(980)$ resonances. In the Appendices, we show the angle 
integrated asymmetry as well as the derivation of kinematical factors.

\section{CPT invariance in a weak decay} 
\label{secCPT}

We follow closely Refs.~\cite{Marshak,Branco,BedPRD14} to introduce our notation 
and the  CPT constraint in  $B$ meson decays. We start by reminding that the weak and 
strong Hamiltonians conserve CPT, namely,
\begin{equation}
({\mathcal 
{CPT}})^{-1}\,H_w\,\,\left(H_s\right)\,{\mathcal {CPT} }= H_w \, \left(H_s\right) \ .
\end{equation}
The hadron weak decay amplitude is computed from the matrix element  $\langle 
\lambda_{out}|H_w|h\rangle$, where the distorted state $\lambda_{out}$ has the effect of 
the hadronic strong force due to  FSI. The requirement of CPT invariance on the decay 
amplitude is
\begin{eqnarray}
\langle \lambda_{out}|H_w|h\rangle
= \chi_h\chi_\lambda\langle  \overline \lambda_{in} |H_w| \overline h  \rangle^*\, ,
\label{cpt3}
\end{eqnarray} 
where we used that the hadron state $|h\rangle$ transforms under CPT as ${\mathcal {CPT} 
}\,|h\rangle=\chi\langle \overline h|$, where $\overline h$ is the charge conjugate 
state and $\chi$ a phase. 

The completeness relation of the strongly interacting states, eigenstates of $H_s$, and 
the hermiticity of $H_w$ implies that
\begin{eqnarray}
\langle \lambda_{out}|H_w|h\rangle
=\chi_h\chi_\lambda\sum_{\overline\lambda^\prime} 
S_{\overline\lambda^\prime,\overline\lambda}
\langle \overline\lambda^{\prime}_{out} |H_w| \overline h  \rangle^*,
\label{cpt4}
\end{eqnarray}
where
\begin{eqnarray}
S_{\overline\lambda^\prime,\overline\lambda}=\langle \overline 
\lambda^\prime_{out}|\overline \lambda_{in}\rangle=S_{\lambda^\prime,\lambda}
\end{eqnarray} 
defines the matrix elements of the S-matrix. The sum of the partial widths of the hadron 
decay channels and the correspondent sum for the charge conjugate  are identical, i.e. the 
mean life of the particle and its conjugate are equal, which is a consequence of 
Eq.~(\ref{cpt4}) and the hermiticity of $H_w$ (see Ref.~\cite{BedPRD14}):
\begin{eqnarray}
\sum_\lambda|\langle \lambda_{out}|H_w|h\rangle|^2
=\sum_{\overline\lambda } 
|\langle \overline\lambda_{out} |H_w| \overline h \rangle|^2 .\,\,\,\,\,\,\,\,
\label{cpt5}
\end{eqnarray} 

The CP-violating phase enters linearly at lowest order in the hadron decay amplitude as 
suggested by the BSS mechanism~\cite{BSS}. In general, the decay amplitude can be written 
as 
\begin{eqnarray}
\mathcal{A}^{\pm} = A_\lambda + B_\lambda e^{\pm i\gamma}, 
\end{eqnarray}
where $A_\lambda$ and $B_\lambda$ are complex amplitudes invariant under CP, containing 
the strongly interacting final-state channel, i.e., 
\begin{eqnarray}
\mathcal{A}^{-}=\langle\lambda_{out} |H_w| h \rangle,
\end{eqnarray}
and 
\begin{eqnarray}
\mathcal{A}^{+}=\langle \overline \lambda_{out} |H_w|  \overline h \rangle.
\end{eqnarray}
The only change due to the CP transformation is the sign multiplying the weak 
phase $\gamma$. The CPT condition depicted by Eq.~(\ref{cpt5}) gives
\begin{eqnarray}
\sum_\lambda \Gamma(A^-_\lambda)  = \sum_{\overline \lambda} \Gamma(A^+_{\overline 
\lambda} ) \ , 
\label{cp5} 
\end{eqnarray}
where the subindex labels the final state channels, summed up in the kinematically 
allowed phase-space. The decay amplitude  written in terms of the CPT constraint 
(\ref{cpt4}), and considering the CP violating amplitudes for the hadron and its charge 
conjugate is
\begin{eqnarray}
A_\lambda+e^{\mp i \,\gamma}B_\lambda
=\chi_h\chi_\lambda\sum_{\lambda^\prime} S_{\lambda^\prime,\lambda}
\left(A _{\lambda^\prime}+e^{\pm i\,\gamma}B_{\lambda^\prime}\right)^*.
\label{cp11}
\end{eqnarray} 
Note that this equation imposes a relation between $A_\lambda$ or $B_\lambda$ with their 
respective complex conjugates.

\section{CP asymmetry and FSI in leading order} 
\label{sect-CPA}

The S-matrix is given in terms of the scattering amplitude $t _{\lambda^\prime,\lambda}$, 
namely,
\begin{eqnarray}
S_{\lambda^\prime,\lambda}=\delta_{\lambda^\prime,\lambda}+i \, t 
_{\lambda^\prime,\lambda},
\end{eqnarray}
which turns Eq.~(\ref{cp11}) into
\begin{eqnarray}
A_\lambda+e^{\mp i \,\gamma}B_\lambda
&=&\chi_h\chi_\lambda \left(A _{\lambda}+e^{\pm i\,\gamma}B_{\lambda}\right)^*
\nonumber\\
&+& 
i \,\chi_h\chi_\lambda \sum_{\lambda^\prime} t_{\lambda^\prime,\lambda}
\left(A _{\lambda^\prime}+e^{\pm i\,\gamma}B_{\lambda^\prime}\right)^*.\quad
\label{cp11-1}
\end{eqnarray}

We now consider effect of the FSI in leading order in the decay amplitudes $A_\lambda$ 
and $B_\lambda$. For that purpose, let us  introduce decay amplitudes computed without the 
effect of the final state interaction,
\begin{eqnarray}
\langle \lambda_0|H_w|h\rangle=A_{0\lambda} + e^{-i\gamma}B_{0\lambda},
\end{eqnarray}
and
\begin{eqnarray}
\langle \overline \lambda_0|H_w|\overline h\rangle=A_{0\lambda} + 
e^{+i\gamma}B_{0\lambda} ,
\end{eqnarray}
where $|\lambda_0\rangle$ and  $|\overline\lambda_0\rangle$ are the mesonic noninteracting 
charge conjugate states, i.e., without the distortion of the strong hadronic interaction. 
The terms $A_{0\lambda}$ and $B_{0\lambda} $ can be in principle associated with the tree 
and penguin amplitudes in the BSS model~\cite{BSS}. 

The leading order (LO) effect due to  FSI in the decay amplitude  is obtained by substituting 
$A_{\lambda}\to A_{0\lambda} $  and $B_{\lambda}\to B_{0\lambda} $   in the 
right-hand side of Eq.~(\ref{cp11-1}).  Considering, the assumption of the CPT invariance 
of the weak Hamiltonian one has also that:
\begin{eqnarray}
\langle \lambda_{0}|H_w|h\rangle
= \chi_h\chi_\lambda\langle  \overline \lambda_{0} |H_w| \overline h  \rangle^*\, ,
\label{cpt3a}
\end{eqnarray} 
which implies in the following relation for the partonic amplitudes
\begin{eqnarray}
A_{0\lambda}=\chi_h\chi_\lambda A_{0\lambda}^*
\end{eqnarray} 
and
\begin{eqnarray}
B_{0\lambda}=\chi_h\chi_\lambda B_{0\lambda}^*,
\label{ab01} 
\end{eqnarray} 
and therefore up to the leading order in $t_{\lambda^\prime,\lambda}$, Eq.~(\ref{cp11-1}) 
reduces to
\begin{eqnarray}
\mathcal{A}^\pm_{LO} &=& A_{0\lambda} + e^{\pm i\gamma}B_{0\lambda}
\nonumber\\
&+& i\sum_{\lambda^\prime}t_{\lambda^\prime,\lambda}\left(A_{0\lambda^\prime} + e^{\pm 
i\gamma}B_{0\lambda^\prime}\right),
\label{cp24}
\end{eqnarray}
which is equivalent to the one provided in Refs.~\cite{wolfenstein,bigibook}.

In lowest order the scattering amplitude is restricted only to two-body terms, therefore 
it is useful to decompose the  source decay amplitudes $A_{0\lambda}$, $B_{0\lambda}$ and 
the $t_{\lambda^\prime,\lambda}$ matrix in angular momentum states $J$ of each pair in 
the outgoing channel. The angular momentum decomposition will allow to identify and 
introduce the meson resonances, like the $\rho(770)$ or $f_0(980)$ in the $\pi\pi$ channel. This 
leading order decay amplitude (\ref{cp24}) decomposed in $J$ is written as,
\begin{eqnarray}
\mathcal{A}^\pm_{LO}&=&\sum_J\left(A^J_{0\lambda}+e^{\pm i\gamma}B^J_{0\lambda} 
\right)
\nonumber \\
&+& i\sum_{\lambda^\prime,J}t^J_{\lambda^\prime,\lambda}
\left(A^J _{0\lambda^\prime}
+e^{\pm i\gamma}B^J_{0\lambda^\prime}\right),
\label{cp24-1}
\end{eqnarray} 
where one should note that $\lambda\,(\lambda^\prime)$ refers to two-body channels in the 
final hadronic state including all other dependences on the quantum numbers and in the energy-momentum 
of the spectator hadrons. By using this expression, the CP asymmetry can be written as
\begin{eqnarray}
&\Delta\Gamma_\lambda& = \Gamma\left(h\to \lambda\right)-\Gamma(\overline h\to 
\overline\lambda)
\nonumber \\
&=&4(\sin\gamma) \, \sum_{J\,J^\prime}\mbox{Im}\left\{ 
\left(B^{J}_{0\lambda}\right)^*A^{J^\prime}_{0\lambda} \right.
\nonumber \\
&+& \left.
i\sum_{\lambda^\prime}\left[\left(B^{J}_{0\lambda}\right)^*t^{J^\prime}_{
\lambda^\prime , \lambda}\,A^{J^\prime}_{0\lambda^\prime} - 
\left(B^{J^\prime}_{0\lambda^\prime}\, t^{J^\prime}_{\lambda^\prime,\lambda}\right)^*
A^J_{0\lambda}\right]\right\},\qquad
\label{cp26}
\end{eqnarray}
where the  $\lambda^\prime$ represents each  state coupled by the strong interaction to 
the  decay channel $\lambda$. We just remind that the second and third terms in the 
right-hand side of Eq.~(\ref{cp26}) can be associated to the ``compound'' CP 
asymmetry~\cite{Soni2005}. These two terms cancel each other when summed in all channels 
$\lambda$ and integrated over the phase-space, that leads to the CPT condition expressed 
by Eq.~(\ref{cp5}), once the source term in (\ref{cp26}) satisfies
\begin{eqnarray}
\sum_{\lambda \,J}\mbox{Im} \left[\left(B^{J}_{0\lambda}\right)^*A^{J}_{0\lambda}\right] = 0,
\label{cp22}
\end{eqnarray}
which is a consequence of the CPT constraint at the microscopic level, e.g., as 
expressed by the tree and penguin amplitudes in the BSS model, that  should be valid when 
FSI is turned off in Eq.~(\ref{cpt3}). This term was neglected by Wolfenstein, which 
corresponds to the trivial solution of Eq.~(\ref{cp22}), assuming that the phase
difference between the two CP-conserving  amplitudes is zero for all decay channels. 

To be complete and detailing the notation of Ref.~\cite{BedPRD14} by including the 
two-particle angular momentum states $J$, we show that the second term in 
Eq.~(\ref{cp26}),
 \begin{eqnarray}
\sum_\lambda\Delta \Gamma^{FSI}_\lambda &=& 4(\sin\gamma)
\sum_{\lambda\,\lambda^\prime\,J}\mbox{Re}\left[\left(B^{J}_{0\lambda}\right)^*t^{J}_{
\lambda^\prime,\lambda}\,A^{J}
_{0\lambda} \right. 
\nonumber \\
&-& \left. \left(B^{J}_{0\lambda}\, t^{J}_{\lambda,\lambda}\right)^*A^J_{0\lambda}\right],
\label{cp26d}
\end{eqnarray}
also satisfies the CPT condition, namely, this quantity vanishes, which is easily verified 
by using Eqs.~(\ref{ab01}) as
\begin{eqnarray}
\sum_\lambda\Delta \Gamma^{FSI}_\lambda &=& 4(\sin\gamma)\times
\nonumber\\
&\times& \sum_{\lambda^\prime\lambda\, J}\mbox{Re}\left[\chi_h\chi_{\lambda J} 
\left(B^J_{0\lambda}\right)^*t^J_{\lambda^\prime,\lambda}\left(A^J_{0\lambda^\prime}
\right)^* \right.
\nonumber \\
&-& \left. \chi^*_h\chi^*_{\lambda^\prime J} 
B^J_{0\lambda^\prime}\left(t^J_{\lambda^\prime,\lambda}\right)^*A^J_{0\lambda}\right] =0.
\label{cp26d2}
\end{eqnarray} 
 
The vanishing of Eq.~(\ref{cp26d2}) is due to the symmetry of 
$t^J_{\lambda,\lambda^\prime}=t^J_{\lambda^\prime,\lambda}$, and the fact that  
$\chi_{\lambda J} =\chi_{\lambda^\prime J} $, i.e., the strong interaction does not mix 
different $CP$ eigenstates. Therefore, by taking into account Eqs.~(\ref{cp22}) and 
(\ref{cp26d2}), one has that the CPT constraint
\begin{eqnarray}
\sum_\lambda\Delta\Gamma_\lambda &=& 4(\sin\gamma)\sum_{\lambda \,J}\mbox{Im} 
\left[\left(B^{J}_{0\lambda}\right)^*A^{J}_{0\lambda}\right]
\nonumber\\
&+&\sum_\lambda\Delta \Gamma^{FSI}_\lambda=0, 
\label{cpt10}
\end{eqnarray}
is fulfilled in leading order of the interaction.

\section{Resonant channels and CPT}
\label{sec:rescpt}

In the case that the channel $\lambda$ contains also the formation of a resonance in the 
partonic process, namely, $B\to \pi\rho$, the amplitudes $A_{0\lambda}$ and $B_{0\lambda}$ 
can be separated in the following two parts, $A^J_{0\lambda}=A^J_{0\lambda NR}+\sum_R 
A^J_{0\lambda R}$, and $B^J_{0\lambda}=B^J_{0\lambda NR}+\sum_R B^J_{0\lambda R}$, where 
the subindex $R$ and $NR$ mean resonant and non resonant channels. Therefore, the decay 
amplitude in Eq.~(\ref{cp24-1}) is rewritten as
\begin{eqnarray}
\mathcal{A}^\pm_{LO} &=& \sum_J\left[\sum_R A^J_{0\lambda R}+A^J_{0\lambda NR}+ \right. 
\nonumber \\
&+& \left. e^{\pm i\gamma}\left(\sum_R B^J_{0\lambda R}+B^J_{0\lambda NR}\right) \right] 
\nonumber \\ 
&+& i\sum_{\lambda^\prime,J}t^J_{\lambda^\prime,\lambda} \left[\sum_R 
A^J_{0\lambda^\prime R}+A^J_{0\lambda^\prime NR} \right.
\nonumber \\
&+& \left. e^{\pm i\gamma}\left(\sum_R B^J_{0\lambda^\prime R}+B^J_{0\lambda^\prime 
NR}\right) \right].
\label{cp24-2}
\end{eqnarray}

The resonant source terms $A^J_{0\lambda R}$ and $B^J_{0\lambda R}$ should be interpreted 
as bare amplitudes, where at the resonance decay vertex, the two-hadron rescattering 
process is not yet included. The Breit-Wigner amplitudes for each term are identified 
according to
\begin{eqnarray}
(1\,+\,i\, t^J_{\lambda\lambda})A^J_{0\lambda R} \to a_0^R 
F_{R\,\lambda}^{BW}P_J(\cos\theta)
\end{eqnarray}
and
\begin{eqnarray}
(1\,+\,i\, t^J_{\lambda\lambda})B^J_{0\lambda R}\to b_{0\lambda}^R 
F_{R\,\lambda}^{BW}P_J(\cos\theta)  \label{ress-1},\quad
\end{eqnarray}
where $J$ is the spin of the resonance decaying to two spin zero particles and 
$P_J(\cos\theta)$ is the Legendre polynomial and $\theta$ is the helicity angle between 
the equally charge particles in the Gottfried-Jackson frame. We will give the 
representation of this angle for the \mbox{$B^+\to\pi^+\pi^+\pi^-$} decay in 
Fig.~\ref{b-pipipi-theta} of the next section.
 
After substituting (\ref{ress-1}) in (\ref{cp24-2}), we get that
\begin{eqnarray}
\mathcal{A}^\pm_{LO} &=& \sum_{J\, R}\left(a_{0\lambda}^R +e^{\pm i\gamma} b_0^R\right) 
F_{R\,\lambda}^{BW}P_J(\cos\theta) 
\nonumber \\
&+& \sum_J \left(A^J_{0\lambda NR}+e^{\pm i\gamma}B^J_{0\lambda NR}\right) 
\nonumber \\ 
&+& i\sum_{\lambda^\prime,J}t^J_{\lambda^\prime,\lambda}\left(A^J_{0\lambda^\prime NR} 
+e^{\pm i\gamma}B^J_{0\lambda^\prime NR} \right),
\label{cp24-3}
\end{eqnarray}
where the first and second terms in the right-hand side is the isobar model for the 
decay. The second term is the source term for the final state channel, and the third one 
includes the hadronic interaction among the two of the hadrons with angular momentum $J$. 
We should clarify that Eq.~(\ref{cp24-3}) includes the interaction in the resonance 
region as the pair of hadrons has a probability to be formed directly from the partonic 
process.

\begin{widetext}
The CP asymmetry from Eq.~(\ref{cp24-3}) can be cast in the following form
\begin{eqnarray}
\Delta \Gamma_\lambda &=& \Gamma\left(h\to \lambda\right)-\Gamma(\overline h\to 
\overline\lambda) 
\nonumber \\ 
&=& 4(\sin\gamma) \, \sum_{J\,J^\prime}\mbox{Im}\left\{  \left(
\sum_{R}b_{0\lambda}^R F_{R\lambda}^{BW}P_J(\cos\theta) + B^J_{0\lambda NR}\right)^* 
\left(\sum_{R^\prime}a_{0\lambda}^{R^\prime}F_{R^\prime\lambda}^{BW}P_{J^\prime}
(\cos\theta) + A^{J^\prime}_{0\lambda NR}\right) \right. 
\nonumber \\ 
&+& \,i \sum_{\lambda^\prime}\left(\sum_{R}b_{0\lambda}^R 
F_{R\lambda}^{BW}P_J(\cos\theta) + B^J_{0\lambda NR}\right)^*\, 
t^{J^\prime}_{\lambda^\prime,\lambda}\, 
\left(\sum_{R^\prime}a_{0\lambda^\prime}^{R^\prime}F_{R^\prime\lambda^\prime}^{BW}P_{
J^\prime}(\cos\theta) + A^{J^\prime}_{0\lambda^\prime NR}\right) 
\nonumber \\ 
&-&\,\left. i \sum_{\lambda^\prime}\left(\sum_{R^\prime}b_{0\lambda^\prime}^{R^\prime} 
F_{R^\prime\lambda^\prime}^{BW}P_{J^\prime}(\cos\theta) + B^{J^\prime}_{0\lambda^\prime 
NR}\right)^*\, \left[t^{J^\prime}_{\lambda^\prime,\lambda}\right]^*\, \left(
\sum_{ R}a_{0\lambda}^{R} F_{R\lambda}^{BW}P_{J}(\cos\theta) 
+A^{J}_{0\lambda NR}\right) \right\} .
\label{cp27b}
\end{eqnarray}
\end{widetext}

It has to be understood that the subindex $\lambda$ also includes  different kinematical 
regions of the three-body channel. Since we have introduced the Breit-Wigner amplitudes 
in the decay amplitude, the CPT constraint has to be checked in the actual fit, i.e., if  
$\sum_\lambda\Delta \Gamma_\lambda=0$ is satisfied when one takes into account the 
integration over the phase-space besides the sum over all decay channels in the sum of 
$\lambda$. Indeed, in our fitting procedure, we will keep only terms that, within our 
limited model, satisfy the CPT constraint.

\section{Interfering resonant and non resonant amplitudes}
\label{sec:interf}
   
We present a simple example to explore the asymmetry formula (\ref{cp27b}), considering 
the resonant, non-resonant source terms and the contribution from the coupling between 
two strongly interacting channels, namely $\pi\pi$ and $KK$. We use the vector and scalar 
resonances, the $\rho(770)$ and $f_0(980)$ ones, for instance, interfering with a non 
resonant amplitude and the term carrying the strong interaction transition amplitude 
between the coupled channels. This illustrates exactly the $B^{\pm}\to 
\pi^{\pm}\pi^+\pi^-$ decay case at low invariant $\pi^+\pi^-$ mass. Also there are 
$B^{\pm}\to K^{\pm}\pi^+\pi^-$ data~\cite{bellekpipi,babarkpipi} previous to the CP 
asymmetry observation by LHCb collaboration.

First, let us remind that, in a general way, the Breit-Wigner excitation curve for a 
resonance $R$ reads 
\begin{eqnarray}
F^{\rm BW}_{R} (s) = \frac{1}{m^2_{R} - s - i 
m_{R}\Gamma_{R}(s)},
\label{BW}
\end{eqnarray}
with $m_R$ being the resonance mass, and
\begin{eqnarray}
\Gamma_{R}(s) = 
\frac{\left(\frac{s}{4}-m_\pi^2\right)^{1/2}m_R\Gamma_R^\prime}{\left(\frac{m_R^2}{4}
-m_\pi^2\right)^ {1/2}s^{1/2}},
\label{ress}
\end{eqnarray}
denoting the energy dependent relativistic width. For the pion mass, we adopted 
$m_\pi=0.138$~GeV, degenerated for the negative and positive charged particles. Here, 
we consider the resonance decay in the $\pi\pi$ channel.

The real and imaginary parts of $F^{\rm BW}_{R}(s)$ are given, respectively, by
\begin{eqnarray}
{\rm Re}\left[F^{\rm BW}_{R}\right] = \frac{m^2_{R}-s}{(m^2_{R} - s)^2+
m_{R}^2\Gamma_{R}(s)^2 }, 
\end{eqnarray}
and
\begin{eqnarray}
{\rm Im}\left[F^{\rm BW}_{R}\right] = \frac{m_{R}\Gamma_{R}(s)}{(m^2_{R} - s)^2+ 
m_{R}^2\Gamma_{R}(s)^2 }.
\label{real}
\end{eqnarray}
The square modulus is
\begin{eqnarray}
{\rm }|F^{\rm BW}_{R}|^2(s) = \frac{1}{(m^2_{R} - s)^2+m_{R}^2\Gamma_{R}(s)^2 }. 
\label{modulus}
\end{eqnarray}

\begin{figure}[!htb]
\centering
\includegraphics[scale=0.69]{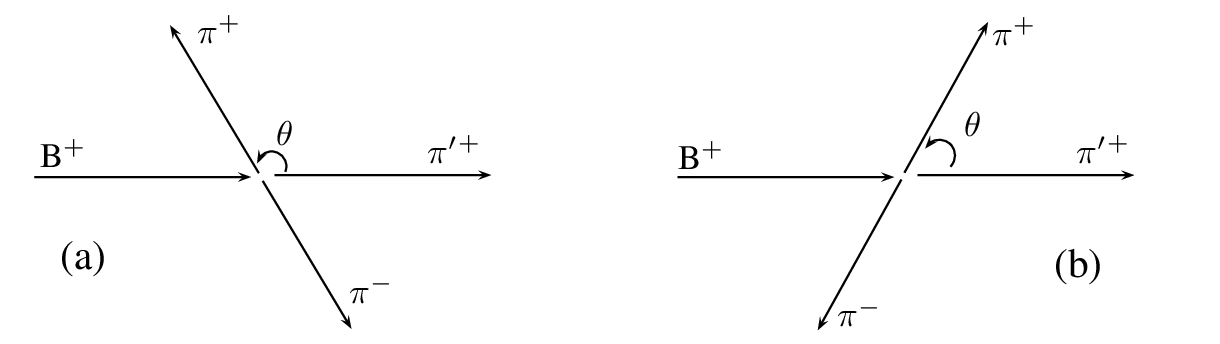}
\caption{\mbox{$B^+\to\pi^+\pi^+\pi^-$} decay with $\pi'^+$ being the bachelor particle.
(a):~$\cos\theta<0$ ($\theta>\frac{\pi}{2}$). (b):~$\cos\theta>0$ 
($\theta<\frac{\pi}{2}$).}
\label{b-pipipi-theta}
\end{figure}

The amplitudes for \mbox{$B^\pm\to\pi^\pm\pi^+\pi^-$} or \mbox{$B^\pm\to 
K^\pm\pi^+\pi^-$} decays, taking into account the $\rho(770)$ and $f_0(980)$ resonances 
interfering with a constant non resonant amplitude, can be written as
\begin{eqnarray}
\mathcal{A}^\pm_{0\lambda} &=& a_0^\rho F^{\rm BW}_{\rho}k(s)\cos\theta+a_0^fF^{\rm BW}_f
+ \frac{a^{nr}_{0\lambda} + b^{nr}_{0\lambda}e^{\pm i\gamma}} 
{1+\frac{s}{\Lambda^2_\lambda}} 
\nonumber \\
&+& [b_0^\rho F^{\rm BW}_{\rho}k(s)\cos\theta + b_0^fF^{\rm BW}_f]e^{\pm i\gamma}, \qquad
\label{ampgeral}
\end{eqnarray}
where the kinematical factor $k(s)=\sqrt{1 - \frac{4m_\pi^2}{s}}$ is included in the 
amplitude of the $\rho(770)$ vector resonance, to take into account the threshold 
behavior of the decay amplitude in a $p$-wave. The angle $\theta$ is defined as the angle 
between the bachelor and the equally charged interacting particle. See this definition in 
the \mbox{$B^+\to\pi^+\pi^+\pi^-$} decay illustrated in Fig.~\ref{b-pipipi-theta}. Here, 
$\cos \theta$ is associated to the spin $1$ of the $\rho$ resonance and varies from $-1$ 
to $+1$ along the phase space. 

The form factor $\left(1+\frac{s}{\Lambda^2_\lambda}\right)^{-1}$ is included in the non 
resonant amplitude in order to parametrize the dependence on the square mass of the pair 
that is brought by the source terms, either for the $\pi\pi$ or $KK$ systems. It comes 
from the partonic decay amplitude that produces the three-meson final state, in which the 
relative momentum between the pair of mesons is distributed among the quarks in the 
momentum loop within  the microscopic amplitude, e. g., the tree and penguin diagrams, 
and probe the internal structure of the mesons involved in the initial and final states. 
In general, the non resonant amplitude should depend on the two kinematically independent 
Mandelstam variables, which for simplicity, we choose one of them. The suggestion is that 
the microscopic process gives raise to the momentum dependence in a power-law form 
reflecting the hard momentum structure of the mesons involved in the decay~\cite{ref8}. 
For example, inspecting the tree diagram, one observes that the decay of the $b-$quark in 
the rest frame, produces a fast light quark and a pion back to back. The light quark has 
to share its momentum with a slow antiquark, which is happens by the gluon exchange and 
is damped by a momentum power-law form. A similar reasoning can be applied to the penguin 
diagram. On the other hand, if factorization can be proved, the heavy to light 
generalized transition form factors~\cite{meissner} like a heavy meson into $K\pi/\pi\pi$, 
for instance, will also play a role in CPV studies.

An alternative to parametrize the decay amplitude, convenient for Monte-Carlo 
simulations, is to write Eq.~(\ref{ampgeral}) as
\begin{eqnarray}
\mathcal{A}_{0\lambda}^\pm &=& a^{\rho}_\pm e^{i \delta^{\rho}_\pm}F^{\rm BW}_{\rho}k(s) 
\cos\theta + a^{f}_\pm e^{i \delta^{f}_\pm}F^{\rm 
BW}_{f} 
+\frac{a^{nr}_{\pm\lambda}e^{i\delta^{nr}_{\pm\lambda}}}{1+\frac{s}{\Lambda^2_\lambda}},
\nonumber\\
\label{ampgeral-mcarlo}
\end{eqnarray}
where $\delta^\rho_{\pm}$ and $\delta^f_{\pm}$ contain both the fixed weak and strong 
phases, with the Breit-Wigner functions introducing additional mass dependent strong 
phases as sketched above. The phase $\delta^{nr}_{\pm}$ comes from the partonic amplitude 
producing the three-body final state, excluding the strong phase from the rescattering  
process. The relation between the parameters is $a^{\rho}_\pm e^{i \delta^{\rho}_\pm} = 
a_0^\rho + b_0^\rho e^{\pm i\gamma}$, $a^f_\pm e^{i \delta^f_\pm} = a_0^f + b_0^f e^{\pm 
i\gamma}$, and $a^{nr}_{\pm\lambda}e^{i \delta^{nr}_{\pm\lambda}} = a^{nr}_{0\lambda} + 
b^{nr}_{0\lambda}e^{\pm i\gamma}$.

By comparing the first term in the r.h.s. of Eqs.~(\ref{cp27b}) and (\ref{ampgeral}), it 
is possible to make the following identifications, namely, 
\begin{eqnarray}
A_{0\lambda R}&=&a_0^\rho F^{\rm BW}_{\rho}k(s)\cos\theta + a_0^fF^{\rm BW}_f,
\nonumber \\
B_{0\lambda R}&=&b_0^\rho F^{\rm BW}_{\rho}k(s)\cos\theta + b_0^fF^{\rm BW}_f,
\nonumber \\
A_{0\lambda NR}&=&\frac{a^{nr}_{0\lambda}}{1+\frac{s}{\Lambda^2_\lambda}},
\nonumber \\
B_{0\lambda NR}&=&\frac{b^{nr}_{0\lambda}}{1+\frac{s}{\Lambda^2_\lambda}},
\end{eqnarray}
and thus, rewrite Eq.~(\ref{cp27b}) for the decay channels $\lambda\,=\,\pi\pi\pi$ or 
$K\pi\pi$ as
\begin{widetext}
\begin{eqnarray}
\Delta\Gamma_\lambda &=&  4(\sin\gamma)\mbox{Im}\Bigg[\left(b_0^\rho F^{\rm 
BW}_{\rho}k(s)\cos\theta + b_0^f F^{\rm BW}_f + 
\frac{b^{nr}_{0\lambda}}{1+\frac{s}{\Lambda^2_\lambda}}\right)^* 
\left(a_0^\rho F^{\rm BW}_{\rho}k(s)\cos\theta + a_0^f F^{\rm BW}_f + 
\frac{a^{nr}_{0\lambda}}{1+\frac{s}{\Lambda^2_\lambda}}\right) \Bigg] 
\nonumber \\ 
&+& 4(\sin\gamma)\,\mbox{Re}\Bigg\{\sum_{\lambda^\prime}\Bigg[ \left(b_0^\rho F^{\rm 
BW}_{\rho}k(s)\cos\theta + b_0^f F^{\rm BW}_f + 
\frac{b^{nr}_{0\lambda}}{1+\frac{s}{\Lambda^2_\lambda}}\right)^* 
\,t^{J=0}_{\lambda^\prime,\lambda}\,
\frac{a^{nr}_{0\lambda^\prime}}{1+\frac{s}{\Lambda^2_{\lambda^\prime}}}\, 
\nonumber\\ 
&-&\left(\frac{b^{nr}_{0\lambda^\prime}}{1+\frac{s}{\Lambda^2_{\lambda^\prime}}}\,\,t^{J=0
}_{\lambda^\prime,\lambda}\right)^*\,\,
\left(a_0^\rho F^{\rm BW}_{\rho}k(s)\cos\theta + a_0^f F^{\rm BW}_f + 
\frac{a^{nr}_{0\lambda}}{1+\frac{s}{\Lambda^2_\lambda}}\right)\Bigg]\Bigg\} ,
\label{cp27d}
\end{eqnarray}
\end{widetext}
where we used the identity $\mbox{Im}(iz)=\mbox{Re}(z)$ and kept only one term in the 
scattering amplitude for $J=0$ in Eq.~(\ref{cp27b}). The subindex $\lambda$ are 
associated with a position in the phase-space of the $B$ decay and $\lambda^\prime$ is 
equal to $\pi KK$ or $KKK$, in the case of $B\to \pi\pi\pi$ and $B\to K\pi\pi$, 
respectively. The second term in the r.h.s. of Eq.~(\ref{cp27d}) is the compound 
contribution to the CP asymmetry that includes the effect of the final state interaction 
in the non resonant channel. In this example, only the $S$-wave amplitude $KK\to\pi\pi$ 
channel was included and we have considered the interference of this term with the 
resonant ones. We use here the functional form of this term obtained in 
Ref.~\cite{BedPRD14}. 

\subsection{CP asymmetry formula}
\label{sec:cpaformula}

The compound contribution to CPV, namely, the one carrying 
$t^{J=0}_{\lambda^\prime,\lambda}$ in Eq.~(\ref{cp27d}), has as input in our calculations 
the non-diagonal scattering amplitude depicted in the full S-matrix in the isoscalar and 
angular momentum zero $\pi\pi\to KK$ coupled channels, which is written as
\begin{equation}
S=
\begin{bmatrix}
\eta e^{2i\delta_{\pi\pi}}  & 
i\sqrt{1-\eta^2}\,e^{i\left(\delta_{\pi\pi}+\delta_{KK}\right)} \\
i\sqrt{1-\eta^2}\,e^{i\left(\delta_{\pi\pi}+\delta_{KK}\right)} & \eta 
e^{2i\delta_{KK}} 
\end{bmatrix}
\label{s-matrix}
\end{equation}
with the inelasticity parameter $\eta(s)$ and the $\pi\pi$ phase-shift 
$\delta_{\pi\pi}(s)$, given by~\cite{pelaprd05},
\begin{eqnarray}
\eta(s)=1-\left(\epsilon_1\frac{k_2}{s^{1/2}}+\epsilon_2\frac{k_2^2}{s}\right)
\frac{M^{\prime2}-s}{s},\
\end{eqnarray}
with
\begin{eqnarray}
k_2 = \frac{\sqrt{s-4m_K^2}}{2},
\end{eqnarray}
and
\begin{eqnarray}
\delta_{\pi\pi}(s)= \frac{1}{2}\cos^{-1}\Bigg\{\frac{\cot^2[\delta_{\pi\pi}(s)] - 
1}{\cot^2[\delta_{\pi\pi}(s)] + 1}\Bigg\},
\end{eqnarray}
with
\begin{eqnarray}
\cot(\delta_{\pi\pi}) = c_0\frac{(s-M_s^2)(M_f^2 - 
s)}{M_f^2 s^{1/2}}\frac{|k_2|}{k_2^2},\qquad
\end{eqnarray}
respectively. In these expressions, we use $m_K=0.494$~GeV, $M'=1.5$~GeV, 
$M_s=0.92$~GeV, $M_f=1.32$~GeV, $\epsilon_1=2.4$, $\epsilon_2=-5.5$, and $c_0=1.3$, 
according to the parametrization given in Ref.~\cite{pelaprd05}. The off-diagonal term in 
the S-matrix gives the transition amplitude between the $\pi\pi$ and $KK$ channels in the 
isoscalar channel, namely, 
$t^{J=0}_{\lambda^\prime,\lambda}=\sqrt{1-\eta^2}\,e^{i\left(\delta_{\pi\pi} +\delta_{KK}
\right)}\approx \sqrt{1-\eta^2}\,e^{2i\delta_{\pi\pi}},$ where we have made the 
approximation $\delta_{KK}\approx\delta{\pi\pi}$ between $1$ and $1.6$~GeV.

The expression for the CP asymmetry in Eq.~(\ref{cp27d}), should be expanded in a form 
where the unknown parameters are exposed in a simple manner to proceed with the fitting to 
the experimental data. By using the relations 
$\mbox{Im}(z_1z+z_2z^*)=\mbox{Re}(z)\mbox{Im}(z_1+z_2) + 
\mbox{Im}(z)\mbox{Re}(z_1-z_2)$, $\mbox{Re}(z_1^*z_2)=\mbox{Re}(z_1)\mbox{Re}(z_2) + 
\mbox{Im}(z_1)\mbox{Im}(z_2)$, and $\mbox{Im}(z_1^*z_2)=\mbox{Re}(z_1)\mbox{Im}(z_2) - 
\mbox{Im}(z_1)\mbox{Re}(z_2)$, together with Eqs.~(\ref{real})-(\ref{modulus}), one can 
finally write Eq.~(\ref{cp27d}) as
\begin{widetext}
\begin{eqnarray}
&\Delta\Gamma_\lambda& = 
\frac{\mathcal{A}}{\left(1+\frac{s}{\Lambda^2_{\lambda}}\right)^2}
+\frac{\left\{\mathcal{B}\cos[2\delta_{\pi\pi}(s)] + 
\mathcal{B}^\prime\sin[2\delta_{\pi\pi}(s)]\right\}\sqrt{1-\eta^2(s)}}{\left(1+\frac{s}{ 
\Lambda^2_{\lambda}}\right)\left(1+\frac{s}{ \Lambda^2_{\lambda^\prime}}\right)} 
%
+ \mathcal{C}|F^{\rm BW}_\rho(s)|^2k^2(s)\cos^2\theta 
\nonumber \\
&+& |F^{\rm BW}_\rho(s)|^2k(s)\cos\theta
\left\{\frac{\mathcal{D}(m^2_\rho-s)}{1+\frac{s} {\Lambda^2_{\lambda}}} 
+\frac{\mathcal{D}^\prime\sqrt{1-\eta^2(s)}\left\{m_\rho\Gamma_\rho(s)\cos[2\delta_{\pi\pi
}(s)]-(m^2_\rho-s)\sin[2\delta_{\pi\pi}(s)]\right\}}{1+\frac{s}{\Lambda^2_{\lambda^\prime}
}} \right.
\nonumber \\
&+& \left. \frac{\mathcal{E}m_\rho\Gamma_\rho(s)}{1+\frac{s} {\Lambda^2_{\lambda}}} 
+\frac{\mathcal{E}^\prime\sqrt{1-\eta^2(s)}\left\{(m^2_\rho-s)\cos[2\delta_{\pi\pi
}(s)]+m_\rho\Gamma_\rho(s)\sin[2\delta_{\pi\pi}(s)]\right\}}{1+\frac{s}{
\Lambda^2_{\lambda^\prime}}}\right\} 
\nonumber \\
&+&|F^{\rm BW}_{\rho}(s)|^2|F^{\rm BW}_f(s)|^2k(s)\cos\theta\times 
\nonumber\\
&\times&\left\{\mathcal{F}[(m_\rho^2 -s)(m_f^2-s) + m_\rho\Gamma_\rho(s)m_f\Gamma_f(s)] 
+\mathcal{G}[(m_\rho^2 -s)m_f\Gamma_f(s) - m_\rho\Gamma_\rho(s)(m_f^2-s)]\right\} 
\nonumber \\
&+& |F^{\rm BW}_f(s)|^2 
\left\{\frac{\mathcal{H}(m^2_f-s)}{1+\frac{s}{\Lambda^2_{\lambda}}} 
+\frac{\mathcal{H}^\prime\sqrt{1-\eta^2(s)}\left\{m_f\Gamma_f(s)\cos[2\delta_{\pi\pi
}(s)]-(m^2_f-s)\sin[2\delta_{\pi\pi}(s)]\right\}}{1+\frac{s}{\Lambda^2_{\lambda^\prime}
}} \right.
\nonumber \\
&+& \left. \frac{\mathcal{P}m_f\Gamma_f(s)}{1+\frac{s} {\Lambda^2_{\lambda}}} 
+\frac{\mathcal{P}^\prime\sqrt{1-\eta^2(s)}\left\{(m^2_f-s)\cos[2\delta_{\pi\pi
}(s)]+m_f\Gamma_f(s)\sin[2\delta_{\pi\pi}(s)]\right\}}{1+\frac{s}{\Lambda^2_{
\lambda^\prime}}}\right\} + \mathcal{Q}|F^{\rm BW}_f(s)|^2,
\label{dgamacos} 
\end{eqnarray}
\end{widetext}
where $\lambda$ is associated to $\pi^\pm\pi^+\pi^-$ or $K^\pm\pi^+\pi^-$, and 
$\lambda^\prime$ to
$\pi^\pm K^+K^-$ or $K^\pm K^+K^-$. The masses of the resonances are
 $m_\rho=0.775$~GeV, and $m_f=0.975$~GeV~\cite{PDG,E791} . The widths  used in 
Eq.~(\ref{ress}) for $R=\rho,f$ are $\Gamma_\rho^\prime=0.150$~GeV and 
$\Gamma_f^\prime=0.044$~GeV. The parameters of the model can be written in terms of those 
in Eqs.~(\ref{ampgeral}) or (\ref{ampgeral-mcarlo}) as follows,
\begin{eqnarray}
\mathcal{A} = 4(\sin\gamma)\mbox{Im}\left[a^{nr}_{0\lambda}b^{nr*}_{0\lambda}\right] 
= (a^{nr}_{+\lambda})^2 - (a^{nr}_{-\lambda})^2,
\end{eqnarray}
\begin{eqnarray}
\mathcal{B} = 4(\sin\gamma)\mbox{Re}\left[a^{nr}_{0\lambda^\prime}b^{nr*}_{0\lambda}-
a^{nr}_{0\lambda}b^{nr*}_{0\lambda^\prime}\right],
\end{eqnarray}
\begin{eqnarray}
\mathcal{B}^\prime = 
-4(\sin\gamma)\mbox{Im}\left[a^{nr}_{0\lambda^\prime}b^{nr*}_{0\lambda}+
a^{nr}_{0\lambda}b^{nr*}_{0\lambda^\prime}\right],
\end{eqnarray}
\begin{eqnarray}
\mathcal{C}&=& 4(\sin\gamma) \, \mbox{Im}\left[a^\rho_0b_0^{\rho*} \right] = 
(a^{\rho}_+)^2 - (a^{\rho}_-)^2,
\end{eqnarray}
\begin{eqnarray}
&\mathcal{D}&=4(\sin\gamma)\mbox{Im}\left[a_0^\rho b^{nr*}_{0\lambda} + 
a^{nr}_{0\lambda}b_0^{\rho*}\right]
\nonumber \\ &=& 
2[a^{\rho}_+a^{nr}_{+\lambda} \cos(\delta^{\rho}_+ - 
\delta^{nr}_{+\lambda}) 
- a^{\rho}_-a^{nr}_{-\lambda}\cos(\delta^{\rho}_- - \delta^{nr}_{-\lambda})],\qquad
\end{eqnarray}
\begin{eqnarray}
&\mathcal{D}^\prime&=4(\sin\gamma)\mbox{Im}\left[a_0^\rho b^{nr*}_{0\lambda^\prime} 
+ a^{nr}_{0\lambda^\prime}b_0^{\rho*}\right]
\nonumber \\ &=& 
2[a^{\rho}_+a^{nr}_{+\lambda^\prime} \cos(\delta^{\rho}_+ - 
\delta^{nr}_{+\lambda^\prime}) 
-a^{\rho}_-a^{nr}_{-\lambda^\prime}\cos(\delta^{\rho}_- - 
\delta^{nr}_{-\lambda^\prime})],\quad\,\,\,
\end{eqnarray}
\begin{eqnarray}
&\mathcal{E}&= 4(\sin\gamma)\mbox{Re}\left[a_0^\rho b^{nr*}_{0\lambda} - 
a^{nr}_{0\lambda}b_0^{\rho*}\right] 
\nonumber \\ &=&
-2[a^{\rho}_+a^{nr}_{+\lambda} \sin(\delta^{\rho}_{+} - 
\delta^{nr}_{+\lambda}) - a^{\rho}_-a^{nr}_{-\lambda}\sin(\delta^{\rho}_- - 
\delta^{nr}_{-\lambda})],\qquad
\end{eqnarray}
\begin{eqnarray}
&\mathcal{E}^\prime& = 4(\sin\gamma)\mbox{Im}\left[-a_0^\rho 
b^{nr*}_{0\lambda^\prime} + a^{nr}_{0\lambda^\prime}b_0^{\rho*}\right]
\nonumber \\ &=& 
2[a^{\rho}_+a^{nr}_{+\lambda^\prime} \sin(\delta^{\rho}_+ - 
\delta^{nr}_{+\lambda^\prime}) + a^{\rho}_-a^{nr}_{-\lambda^\prime}\sin(\delta^{\rho}_- 
- \delta^{nr}_{-\lambda^\prime})],\quad\,\,\,
\end{eqnarray}
\begin{eqnarray}
\mathcal{F} &=& 4(\sin\gamma)\mbox{Im}\left[a_0^f b^{\rho*}_0 + 
a^{\rho}_0b_0^{f*}\right]
\nonumber\\ &=&
2[a^{\rho}_+a^{f}_+ \cos(\delta^{\rho}_+ 
- \delta^{f}_+) - a^{\rho}_-a^{f}_- \cos(\delta^{\rho}_- - \delta^{f}_-)],\qquad
\end{eqnarray}
\begin{eqnarray}
\mathcal{G} &=& 4(\sin\gamma)\mbox{Re}\left[a_0^f b^{\rho*}_0 - 
a^{\rho}_0b_0^{f*}\right] 
\nonumber \\ &=&
2[a^{\rho}_+a^{f}_+ \sin(\delta^{\rho}_+ - \delta^{f}_+) - a^{\rho}_-a^{f}_- 
\sin(\delta^{\rho}_- - \delta^{f}_-)],\qquad
\end{eqnarray}
\begin{eqnarray}
&\mathcal{H}&=4(\sin\gamma)\mbox{Im}\left[a_0^f b^{nr*}_{0\lambda} + 
a^{nr}_{0\lambda}b_0^{f*}\right]
\nonumber \\ &=&
2[a^{f}_+a^{nr}_{+\lambda} \cos(\delta^{f}_+ - 
\delta^{nr}_{+\lambda}) -a^{f}_-a^{nr}_{-\lambda}\cos(\delta^{f}_- - 
\delta^{nr}_{-\lambda})],\qquad
\end{eqnarray}
\begin{eqnarray}
&\mathcal{H}^\prime&=4(\sin\gamma)\mbox{Im}\left[a_0^f b^{nr*}_{0\lambda^\prime} + 
a^{nr}_{0\lambda^\prime}b_0^{f*}\right]
\nonumber \\ &=& 
2[a^{f}_+a^{nr}_{+\lambda^\prime} \cos(\delta^{f}_+ - 
\delta^{nr}_{+\lambda^\prime})
 - a^{f}_-a^{nr}_{-\lambda^\prime}\cos(\delta^{f}_- - 
\delta^{nr}_{-\lambda^\prime})],\quad\,\,\,
\end{eqnarray}
\begin{eqnarray}
&\mathcal{P}& = 4(\sin\gamma)\mbox{Re}\left[a_0^f b^{nr*}_{0\lambda} - 
a^{nr}_{0\lambda}b_0^{f*}\right]
\nonumber \\ &=&
-2[a^{f}_+a^{nr}_{+\lambda} \sin(\delta^{f}_{+} - \delta^{nr}_{+\lambda}) 
-a^{f}_-a^{nr}_{-\lambda}\sin(\delta^{f}_- - \delta^{nr}_{-\lambda})],\quad\,
\end{eqnarray}
\begin{eqnarray}
&\mathcal{P}^\prime&=4(\sin\gamma)\mbox{Im}\left[-a_0^f b^{nr*}_{0\lambda^\prime} 
+ a^{nr}_{0\lambda^\prime}b_0^{f*}\right]
\nonumber \\ &=& 
2[a^{f}_+a^{nr}_{+\lambda^\prime} \sin(\delta^{f}_+ - 
\delta^{nr}_{+\lambda^\prime}) 
+ a^{f}_-a^{nr}_{-\lambda^\prime}\sin(\delta^{f}_- - 
\delta^{nr}_{-\lambda^\prime})],\quad\,\,\,
\end{eqnarray}
and
\begin{eqnarray}
\mathcal{Q} &=& 4(\sin\gamma)\mbox{Im}\left[a^f_0b^{f*}_0\right] = (a^{f}_+)^2 - 
(a^{f}_-)^2.
\end{eqnarray}

In the previous work of Ref.~\cite{BedPRD14}, we used a different form for the $KK 
\rightarrow \pi\pi$ amplitude, writing it as $|K_\lambda| \cos (\delta_\lambda + 
\delta_{\lambda^\prime} + \Phi_\lambda)$, where $K_\lambda = B^*_{0 \lambda} 
A_{0\lambda^\prime} - B_{0 \lambda} A^*_{0 \lambda^\prime}$ and $\Phi_{\lambda} = - i 
\ln(K_\lambda/|K_\lambda|)$. This form is analogous of that used in Eq.~(\ref{dgamacos}), 
in terms of $\mathcal{B}$ and $\mathcal{B^\prime}$. The phase $\Phi_\lambda$ was chosen to 
be zero that is the same as to take $\mathcal{B^\prime} = 0$, assumed hereafter.

As a remark, we call the reader attention that for the compound contribution, we have 
discarded three-body rescattering effects at the two-loop level since they are small 
compared to the first two-body collision contribution, as suggested by the three-body
model calculations for the $D^\pm\to K^\pm\pi^+\pi^-$ decay, see 
Refs.~\cite{dkpipi1,dkpipi2,dkpipi3}.

\section{Analysis of CP asymmetry terms}
\label{sec:analysis}

In order to be able to compare our model directly with the recent experimental data 
regarding the \mbox{$B^{\pm}\to \pi^{\pm} \pi^+\pi^-$} and \mbox{$B^{\pm}\to K^{\pm} 
\pi^+\pi^-$} decays presented in Ref.~\cite{expnew}, we had to eliminate one variable in 
Eq.~(\ref{dgamacos}), namely, the square mass of the pair presenting the bachelor 
particle. We perform this calculation in Appendix~\ref{app-int}. Here in this section, we 
show the behavior of some terms of the integrated Eq.~(\ref{dgamacos}).

\subsection{Direct CPV term containing the $\mathcal{C}$ parameter} 

\begin{figure}[!htb]
\centering
\includegraphics[scale=0.3]{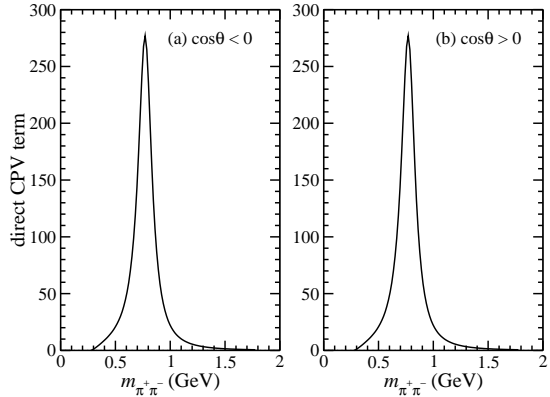}
\caption{Term containing the $\mathcal{C}$ parameter of the integrated 
Eq.~(\ref{dgamacos}), as a function of $m_{\pi^+\pi^-}=\sqrt{s}$ for (a) $\cos\theta<0$ 
and (b) $\cos\theta>0$. In these plots we set $\mathcal{C}=1$.}
\label{dcpv}
\end{figure}

As it was mentioned before, the term of the integrated Eq.~(\ref{dgamacos}) containing 
the $\mathcal{C}$ parameter, is equivalent to the direct CPV induced by the interference 
between the tree and penguin amplitudes with one meson resonance in the final state. In 
this case, the contribution from CPV comes only from the $\rho(770)$ meson. In 
Fig.~\ref{dcpv}, we show the signature of the CPV from this term with the 
usual Breit-Wigner square modulus and with the same sign for both cases, namely, 
$\cos\theta<0$ and $\cos\theta>0$. This term locally violates the CPT constraint and will 
be set to zero in our fittings, by assuming $\mathcal{C}=0$.

\subsection{DCPV term containing the $\mathcal{D}$ parameter} 

\begin{figure}[!htb]
\centering
\includegraphics[scale=0.3]{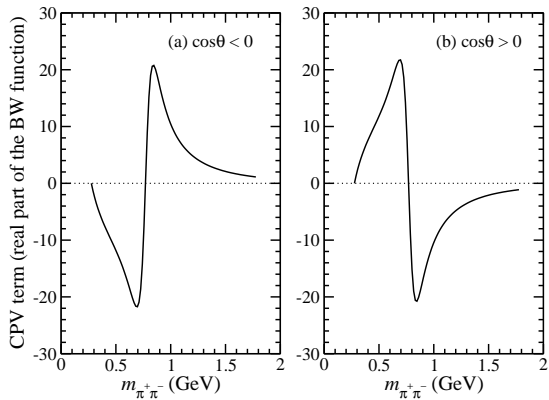}
\caption{Term containing the $\mathcal{D}$ parameter of the integrated 
Eq.~(\ref{dgamacos}), as a function of $m_{\pi^+\pi^-}=\sqrt{s}$ for (a) $\cos\theta<0$ 
and (b) $\cos\theta>0$. Here we have used $\Lambda_\lambda=1.705$~GeV and 
$\mathcal{D}=1$.}
\label{dcpv-real}
\end{figure}

The Dalitz CPV induced by the interference term of the integrated Eq.~(\ref{dgamacos}), 
namely, that containing the $\mathcal{D}$ parameter, is directly related with the real 
part of the Breit-Wigner function. This function presents a clear signature in the mass 
spectrum, namely, a zero and a change in the sign of the CP asymmetry at the central 
value of the $\rho(770)$ mass. Another peculiarity associated to vector mesons resonances 
is one more sign change when the $\cos\theta$ cross the zero around the middle of the 
Dalitz plot. These two facts can be observed in Fig.~\ref{dcpv-real}. Notice that this 
term sums to zero and thus locally satisfies the CPT constraint.

\subsection{DCPV term containing the $\mathcal{E}$ parameter} 

\begin{figure}[!htb]
\centering
\includegraphics[scale=0.3]{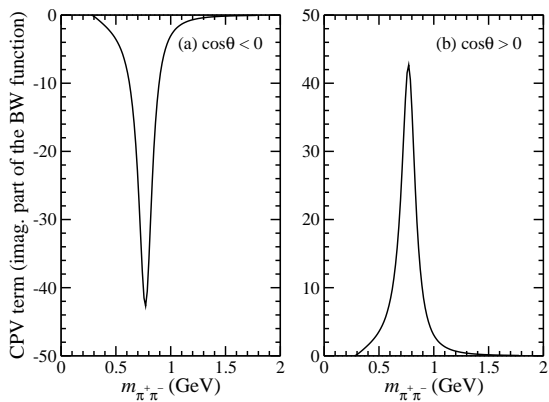}
\caption{Term containing the $\mathcal{E}$ parameter of the integrated 
Eq.~(\ref{dgamacos}), as a function of $m_{\pi^+\pi^-}=\sqrt{s}$ for (a) $\cos\theta<0$ 
and (b) $\cos\theta>0$. Here we have used $\Lambda_\lambda=1.705$~GeV and 
$\mathcal{E}=1$.}
\label{dcpv-imag}
\end{figure}

The term containing the $\mathcal{E}$ parameter in the integrated 
Eq.~(\ref{dgamacos}), associated to the Dalitz CPV induced by the interference between 
the $\rho(770)$ and the non resonant amplitude, is directly related with the imaginary 
part of the Breit-Wigner function of $\rho(770)$. The shape is similar to the those 
presented in Fig.~\ref{dcpv}, with the clear difference that the proportionality with 
$\cos\theta$ changes the sign of the CP asymmetry when $\cos\theta$ pass through zero in 
the middle of the Dalitz plot. Fig.~\ref{dcpv-imag} shows these features. Notice that 
this term sums to zero and thus locally satisfies the CPT constraint.

\subsection{DCPV term containing the $\mathcal{F}$ parameter} 
\begin{figure}[!htb]
\centering
\includegraphics[scale=0.3]{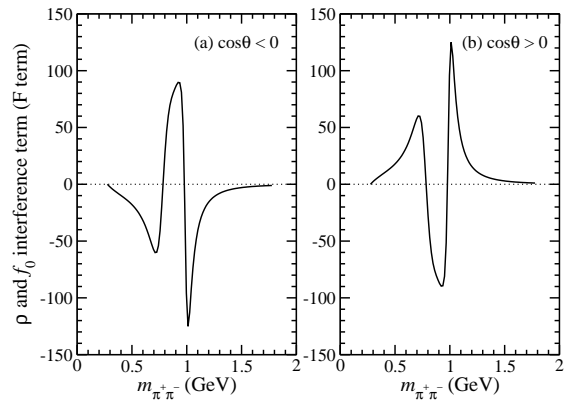}
\caption{Term containing the $\mathcal{F}$ parameter of the integrated 
Eq.~(\ref{dgamacos}), as a function of $m_{\pi^+\pi^-}=\sqrt{s}$ for (a) $\cos\theta<0$ 
and (b) $\cos\theta>0$. Here we set $\mathcal{F}=1$.}
\label{dcpv-f}
\end{figure}

The projection of the difference between $B^-$ and $B^+$ events presents a clear signature 
in the mass spectrum with a zero close to the mass of the vector meson $\rho$, and another 
one close to the mass of the scalar $f_0(980)$. There is also a change of sign in the CP 
asymmetry associated to the $\cos\theta$ passing through zero around the middle of the 
Dalitz plot. These features can be observed in Fig.~\ref{dcpv-f}, where we display the 
term containing the $\mathcal{F}$ parameter of the integrated Eq.~(\ref{dgamacos}). This 
term locally satisfies the CPT constraint.

\subsection{DCPV term containing the $\mathcal{G}$ parameter} 

\begin{figure}[!htb]
\centering
\includegraphics[scale=0.3]{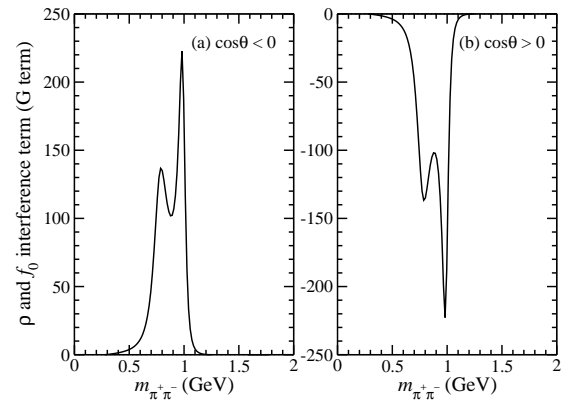}
\caption{Term containing the $\mathcal{G}$ parameter of the integrated 
Eq.~(\ref{dgamacos}), as a function of $m_{\pi^+\pi^-}=\sqrt{s}$ for (a) $\cos\theta<0$ 
and (b) $\cos\theta>0$. Here we set $\mathcal{G}=1$.}
\label{dcpv-g}
\end{figure}

The shape of this term presents the characteristic peaks of the two interfering 
resonances. Due to the direct proportionality to $\cos\theta $, there is a change of sign 
of the CP asymmetry when $\cos\theta$ pass through zero in the middle of the Dalitz plot. 
Fig.~\ref{dcpv-g} shows the features of this term. This term is consistent with the CPT 
constraint locally.

\section{CPV in $B^\pm\to\pi^\pm\pi^+\pi^-$ and $B^\pm\to\pi^\pm K^+K^-$ decays}
\label{sec:cpvfit1}

We determine the relative contributions related to the integrated Eq.~(\ref{dgamacos}) 
by fitting two CP asymmetries distributions presented in Ref.~\cite{expnew}; one for the 
$\pi^+\pi^-$ mass distribution coming from the difference between $B^-$ and $B^+$ in 
$B^{\pm}\to \pi^{\pm} \pi^+\pi^-$ decay, in particular the distribution with 
$\cos\theta>0$. From this fitting we found the parameters and plot the CP asymmetry for 
$\cos\theta<0$. The other fit was performed  for the phase-space integration in the region 
$\cos\theta < 0$ of the $\pi^+\pi^-$ mass distributions in the $B^{\pm}\to K^{\pm} 
\pi^+\pi^-$ decay. From these two fits we get  the value of the  $\pi^+\pi^- \to K^+K^-$ 
parameter $({\mathcal B})$ for each decay channel and plot the asymmetry for the $K^+K^-$ 
distribution in the channels $B^{\pm}\to \pi^{\pm} K^+K^-$ and  $B^{\pm}\to K^{\pm} 
K^+K^-$. 

To perform both fits and the other plots, we use the  $\pi^+\pi^- \to K^+K^-$ amplitude, 
the non resonant component, and the $\rho$ and $f_0(980)$ resonances. We put these 
amplitudes within the isobar model through a coherent sum of them. The biggest source of 
uncertainty from these fits is the error in the $\pi\pi$ phase shift and inelasticity 
parameter. For simplicity we use the central value of Ref.~\cite{pelaprd05}, although 
possible variations within the quoted errors in the parameters could give better results. 
Despite our belief that this model is able to represent much of this data, we are aware 
that it can not explain all the rich CP violation structure observed in these decays. The 
inclusion of other contributions and the symmetrization of the $B^{\pm}\to \pi^{\pm} 
\pi^+\pi^-$ decay amplitude would be necessary to understand in more details the phase 
space of these decays and get better agreement in all CP asymmetry regions presented in 
Ref.~\cite{expnew}. 

To perform the fits we use for the $f_0(980)$ resonance, $m_f=0.975$~GeV and width 
$\Gamma_f^\prime=0.044$~GeV, getting from E791 experiment~\cite{E791}. For the two 
$\Lambda$ parameters, we use $\Lambda_\lambda=\Lambda_{\pi\pi} = 3.0$~GeV and 
$\Lambda_{\lambda^\prime}=\Lambda_{KK} = 4.0$~GeV. It is important to say that we changed 
these values by a factor two without finding an appreciable change in the fitting.

\subsection{$\pi\pi\pi$ channel}\label{sec:Kpipi}

\begin{figure}[!htb]
\centering
{\includegraphics[scale=0.26]{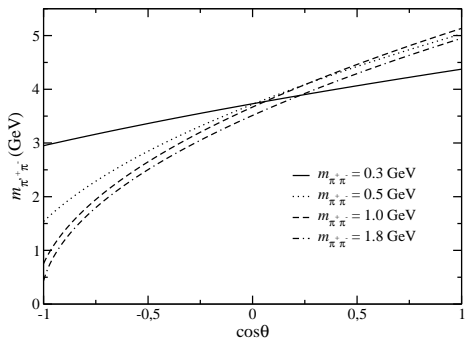}}
\caption{$m_{\pi'^+\pi^-}$ as a function of $\cos\theta$ for some particular values of 
$m_{\pi^+\pi^-}$.}
\label{m32}
\end{figure}

We started by performing the fit of the integrated Eq.~(\ref{dgamacos}) to the 
$\cos\theta>0$ asymmetry distribution of the low $\pi^+\pi^-$ mass projection for the 
$B^{\pm}\to \pi^{\pm} \pi^+\pi^-$ decay. Due the presence of identical particles in final 
state, the interference between symmetrical terms by the exchange of the two identical 
pions must disturb the CP violation pattern mostly for $\cos\theta<0$. We study the 
regions on the Dalitz plot where this interference can be minimum and observe that it 
corresponds to the high $\pi^+\pi^- $ region on the $\cos\theta > 0$ distribution as can 
be seen in Fig.~\ref{m32}. 

This figure provides the mass of the changeless pair of pions formed with the bachelor 
pion. If $\cos\theta<0$ the mass of this pair stays below 3 GeV, and for 
$m_{\pi^{\prime+}\pi^-}\lesssim 1$ GeV, it can be even below 1 GeV, and therefore making 
relevant the interference with the resonances. For $\cos\theta>0$, 
$m_{\pi^{\prime+}\pi^-}> 3$ GeV, minimizing interference effects from the Bose 
symmetrization of the decay amplitude. We intend to perform this study in the future. For 
the moment, we do not consider the symmetrization of the decay amplitude, and therefore we 
do not use the data for the CP asymmetry in the $B^{\pm}\to \pi^{\pm} \pi^+\pi^-$ channel 
for $\cos\theta<0$ to fit the parameters.

\begin{figure}[!htb]
\centering
\psfrag{B --> ppp}[b][b][0.8]{$B^\pm\to\pi^\pm\pi^+\pi^-$}
\psfrag{cost > 0}[b][b][0.8]{$\cos\theta>0$}
\psfrag{B term}[b][b][0.8]{$\mathcal{B}$ term}
\psfrag{D term}[b][b][0.8]{$\mathcal{D}$ term}
\psfrag{E term}[b][b][0.8]{$\mathcal{E}$ term}
\psfrag{G term}[b][b][0.8]{$\mathcal{G}$ term}
{\includegraphics[scale=0.35]{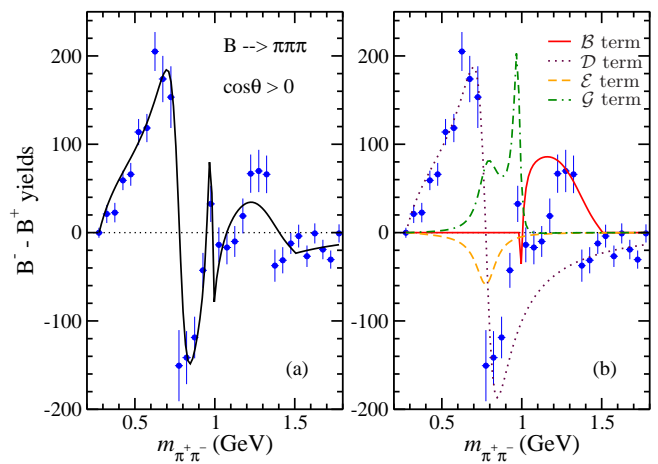}}
\caption{(Color online) CP asymmetry of the $B^\pm \to \pi^\pm\pi^+\pi^-$ decay, 
integrated Eq.~(\ref{dgamacos}), compared with the experimental values (blue points) 
taken from Ref.~\cite{expnew}. Results for $\cos\theta>0$ for (a) total and (b) 
individual contributions.}
\label{b-pipipi-cos-pos}
\end{figure}

The best fit was obtained by using four parameters associated to the $\pi^+\pi^- \to 
K^+K^-$ amplitude ($\mathcal{B}$), the real and imaginary parts of the interference 
between the $\rho$ and the non resonant partonic amplitudes ($\mathcal{D}$ and 
$\mathcal{E}$), and finally the imaginary part of the interference between the $\rho$ and 
the $f_0(980)$ resonances ($\mathcal{G}$). The result is presented in 
Fig.~\ref{b-pipipi-cos-pos}{\color{purple}a}, where we compared our model 
with the experimental data extracted from Fig.~4c of Ref.~\cite{expnew}. The remaining 
parameters, namely, those that locally violate the CPT constraint, and those negligible in 
the fitting procedure, were set to zero.

The individual contributions to the CP asymmetry are shown in 
Fig.~\ref{b-pipipi-cos-pos}{\color{purple}b}. As expected, the $\rho$ meson 
contribution represented by the amplitudes containing the $\mathcal{D}$ and $\mathcal{E}$ 
parameters, is large mainly for the real part of the BW, and the presence of $f_0(980)$ is 
seen only by one of the interfering terms with the $\rho$, namely, that presenting the 
$\mathcal{G}$ parameter. All these terms are locally CPT invariant, namely by integration 
in $\cos \theta$ their contribution to the CP asymmetry vanishes. Between $1$ and 
$1.6$~GeV, the contribution of the $\pi^+\pi^- \to K^+K^-$ amplitude, namely, the 
$\mathcal{B}$ term, shows its importance and as expected~\cite{BedPRD14}, dominates the 
asymmetry in this region. We remind that this part of the asymmetry does not vanishes upon 
integration in $\cos\theta$ and cancels the asymmetry in the $B^{\pm}\to \pi^{\pm} K^+K^-$ 
decay channel.

\begin{figure}[!htb]
\centering
\psfrag{B --> ppp}[b][b][0.8]{$B^\pm\to\pi^\pm\pi^+\pi^-$}
\psfrag{cost < 0}[b][b][0.8]{$\cos\theta<0$}
\psfrag{B term}[b][b][0.8]{$\mathcal{B}$ term}
\psfrag{D term}[b][b][0.8]{$\mathcal{D}$ term}
\psfrag{E term}[b][b][0.8]{$\mathcal{E}$ term}
\psfrag{G term}[b][b][0.8]{$\mathcal{G}$ term}
{\includegraphics[scale=0.35]{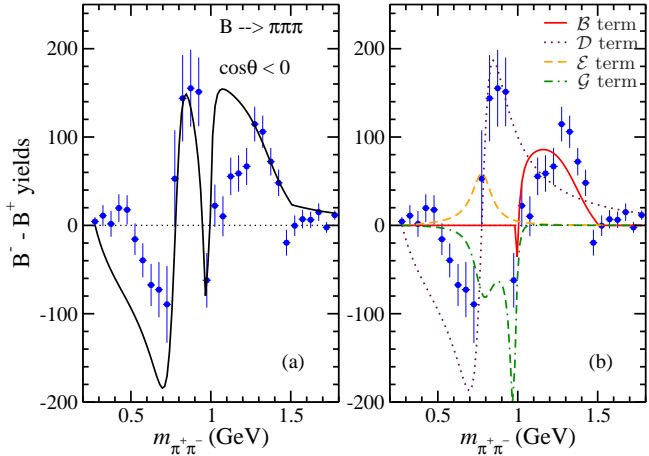}}
\caption{(Color online) CP asymmetry of the $B^\pm \to \pi^\pm\pi^+\pi^-$ decay, 
integrated Eq.~(\ref{dgamacos}), compared with the experimental values (blue points) 
taken from Fig.~4d of Ref.~\cite{expnew}. Results for $\cos\theta<0$ for (a) total and 
(b) individual contributions.}
\label{b-pipipi-cos-neg}
\end{figure}

Looking at the behavior of the experimental $\cos\theta < 0$ distribution in 
Fig.~\ref{b-pipipi-cos-neg}{\color{purple}a}, we can see that much of the features 
observed for the distribution of events for $\cos\theta > 0$ are present in this 
asymmetry. In fact, we plotted together with the experimental points, the amplitudes 
computed with the integrated Eq.~(\ref{dgamacos}) using the fit parameters obtained 
from the $\cos\theta>0$ distribution. The plot in 
Fig.~\ref{b-pipipi-cos-neg}{\color{purple}a} shows a clear departure from the 
experimental data when  at the starting of the $\rho$ mass resonance and also at the 
beginning of the contribution of the $\pi^+\pi^- \to K^+K^-$ amplitude to the asymmetry. 
Assuming that these differences are due to the symmetrization, this result suggests  an 
interference between $\rho$ and $\pi^+\pi^- \to K^+K^-$ amplitude in the crossing 
channels, which is corroborated by Fig.~\ref{m32}, where for the mass region above 
$1$~GeV, the mass of the $\pi^{\prime+}\pi^-$ pair in the crossing channel for 
$\cos\theta\lesssim -0.75$ can be even below $1$ GeV. In addition, we show in 
Fig.~\ref{b-pipipi-cos-neg}{\color{purple}b} the individual contributions to this CP 
asymmetry distribution. 

We still remark to the reader that for the $B^{\pm}\to \pi^{\pm} \pi^+\pi^-$ decay, our 
definition of the angle $\theta$ is  opposite from that presented in 
Ref.~\cite{expnew}. Thus, we had to compare our results of $\cos\theta>0$ 
($\cos\theta<0$) with those of $\cos\theta<0$ ($\cos\theta>0$) of Ref.~\cite{expnew}.

\subsection{$\pi KK$ channel}
\label{sec:pikk}

The expression for the CP asymmetry in the coupled $B^\pm\to\pi^\pm K^+K^-$ channel, 
derived from the general formula in Eq.~(\ref{cp27b}), applied to this specific decay and 
integrated in $\cos\theta$, is given by 
\begin{eqnarray}
\Delta\Gamma(s) &=& -\frac{2\mathcal{A}}{a'(s)\sqrt{s-4m_K^2}\left(1 + 
\frac{s}{\Lambda_{\lambda^\prime}^2}\right)^2}
\nonumber \\
&-& \frac{2\mathcal{B}\sqrt{1-\eta^2(s)}\cos[2\delta_{\pi\pi}(s)]}{a'(s)\sqrt{s-4m_K^2} 
\left(1 + \frac{s}{\Lambda_{\lambda}^2}\right) \left(1 + 
\frac{s}{\Lambda_{\lambda^\prime}^2}\right) },
\label{dgpikk} 
\end{eqnarray}
where the kinematical factors, namely,
\begin{eqnarray}
a'(s)=\frac{1}{(s-4m_K^2)^{1/2}\left[\frac{(M_B^2-m_\pi^2-s)^2}{4s}-m_\pi^2\right]^{1/2}},
\label{aprime}
\end{eqnarray}
and the kaon momentum in the rest frame of the $KK$ subsystem, $\sqrt{s-4m_K^2}$, are now 
taken for the $\pi KK$ system. Furthermore, the integrated decay width in $\cos\theta$ 
from Eq.~(\ref{dgpikk}), becomes exactly opposite in sign to the decay width in the 
$\pi\pi\pi$ channel above the $KK$ threshold, c.f. the first two terms of the 
integrated Eq.~(\ref{dgamacos}).

\begin{figure}[!htb]
\centering
\psfrag{B --> pKK}[b][b][0.8]{$B^\pm\to\pi^\pm K^+K^-$}
{\includegraphics[scale=0.26]{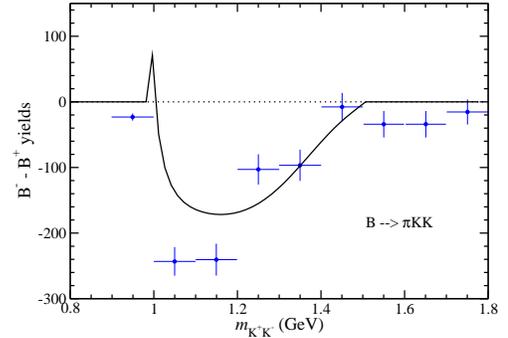}}
\caption{(Color online) CP asymmetry of the \mbox{$B^\pm\to\pi^\pm K^+K^-$} decay, 
Eq.~(\ref{dgpikk}), compared with experimental data (blue points) taken from Fig.~7b of 
Ref.~\cite{expnew}.}
\label{b-pikk}
\end{figure}

The expression in Eq.~(\ref{dgpikk}) is obtained by adding the $\pi^+\pi^- \to K^+K^-$ 
contributions detailed in Figs.~\ref{b-pipipi-cos-pos} and \ref{b-pipipi-cos-neg}, 
since $\mathcal{A}=0$. The plot shown in Fig.~\ref{b-pikk} was done with the 
parameters fitted by the CP asymmetry data in the \mbox{$B^\pm\to\pi^\pm \pi^+\pi^-$} 
decay. The CP violation distribution obtained for the $K^+K^-$ invariant mass from 
$B^{\pm}\to \pi^{\pm} K^+K^-$ decay through Eq.~(\ref{dgpikk}), has opposite sign with 
respect to the correspondent term in the coupled $\pi^\pm\pi^+\pi^-$ channel.

For this decay, the LHCb experiment presented the sum of events with $\cos\theta<0$ and 
$\cos\theta>0$. The comparison of the data with our model is also provided in 
Fig.~\ref{b-pikk}. There is a clear agreement about the shape of the model distribution 
with the experimental data and also a reasonable amount of the number of events related 
with this kind of CP asymmetry. It is also worth to mention that CPT is conserved if we 
sum all CP asymmetry contributions obtained with our approach to $B^{\pm}\to \pi^{\pm} 
\pi^+\pi^-$ and $B^{\pm}\to \pi^{\pm} K^+K^-$ decays, in a region of $\pi^+\pi^-$ and 
$K^+K^-$ invariant mass below $1.6$~GeV.

\section{CPV in $B^\pm\to K^\pm\pi^+\pi^-$ and $B^\pm\to K^\pm K^+K^-$ decays}
\label{sec:cpvfit2}

The number of events observed in these two decay channels are about one order of 
magnitude larger than in the $\pi^\pm\pi^+\pi^-$ and $\pi^\pm K^+K^-$ data~\cite{expnew}. 
This  allows a better fit to the difference between $B^-$ and $B^+$ events. In the 
fitting procedure we use only the $\cos\theta<0$ distribution, because the experimental 
results~\cite{expnew} for $\cos\theta>0$ present a new feature in both decays in the 
region studied in this work. Our simple model does not account for this new behavior. We 
only have a guess, that we already mentioned in our previous paper~\cite{BedPRD14}, 
related with the possible presence of re-scattering coming from  double charm decays. One 
should note that three light-pseudoscalar mesons can, in principle, couple via strong 
interaction with channels  like $D\overline D K$. It seems reasonable to expect that 
$D\overline DK \to KKK$ or $K\pi\pi $ can contribute to the CP asymmetry in regions of 
large two-body invariant mass above the $ D\overline D $ threshold, that is far from the 
$KK$ threshold and above $1.6$~GeV, outside the region discussed in this work but excluded 
from the $\cos\theta<0$ distribution.

\subsection{$K\pi\pi$ channel}
\label{sec:kpipi}

\begin{figure}[!htb]
\centering
\psfrag{B --> Kpp}[b][b][0.8]{$B^\pm\to K^\pm\pi^+\pi^-$}
\psfrag{cost < 0}[b][b][0.8]{$\cos\theta<0$}
\psfrag{B term}[b][b][0.8]{$\mathcal{B}$ term}
\psfrag{D term}[b][b][0.8]{$\mathcal{D}$ term}
\psfrag{F term}[b][b][0.8]{$\mathcal{F}$ term}
\psfrag{G term}[b][b][0.8]{$\mathcal{G}$ term}
\psfrag{D' term}[b][b][0.8]{$\mathcal{D}'$ term}
{\includegraphics[scale=0.35]{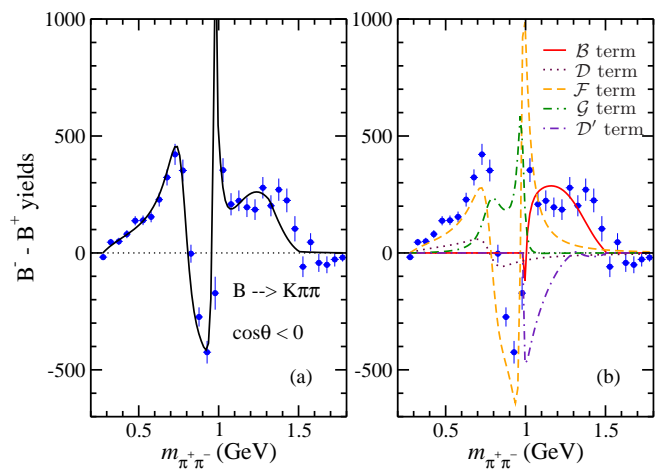}}
\caption{(Color online) CP asymmetry of the $B^\pm \to K^\pm\pi^+\pi^-$ decay, 
integrated Eq.~(\ref{dgamacos}), compared with the experimental values (blue points) 
taken from Fig.~5c of Ref.~\cite{expnew}. Results for $\cos\theta<0$ for (a) total and 
(b) individual contributions.}
\label{b-kpipi-cos-neg}
\end{figure}

Differently to the $B^{\pm}\to \pi^{\pm} \pi^+\pi^-$ decay where the $\rho$ amplitudes 
were dominant, here the $f_0(980)$ has the largest contribution to the $B^{\pm}\to 
K^{\pm} \pi^+\pi^-$ decay, as it is expected by recalling the BSS mechanism applied to 
these decays, which builds in principle the corresponding amplitudes associated with the 
source terms or partonic amplitudes. One possible good fit for the integrated 
Eq.~(\ref{dgamacos}) is obtained by using five parameters associated to the 
$\pi^+\pi^-\to K^+K^-$ amplitude ($\mathcal{B}$), the real parts of the interference 
between $\rho$ resonance with the non resonant partonic ($\mathcal{D}$), and $\pi^+\pi^- 
\to K^+K^-$ amplitudes ($\mathcal{D^\prime}$), and finally the real and imaginary parts of 
the interference between the $\rho$ and the $f_0(980)$ resonances, ($\mathcal{F}$ and 
$\mathcal{G}$). As in the $B^{\pm}\to \pi^{\pm} \pi^+\pi^-$ case, the parameters that 
locally violate the CPT constraint, and those negligible in the fitting procedure were 
set to zero. Fig.~\ref{b-kpipi-cos-neg}{\color{purple}a} shown the best fit we get to 
the $\cos\theta < 0$ distribution. 

In Fig.~\ref{b-kpipi-cos-neg}{\color{purple}b}, we shown the decomposition of each 
component of the fit. Clearly the dominant contribution comes from the real part of the 
interference between $\rho$ and the $f_0(980)$ resonances. The imaginary part of this 
amplitude has a contribution, however, much less important than the real one. The 
$\pi^+\pi^-$ and $ K^+K^-$ amplitude plays an important role in this fit above $1$~GeV 
and also associated with the interference with the $\rho$ resonance. 

Here, we also used $\Lambda_\lambda = 3.0$~GeV and $\Lambda_{\lambda^\prime} 
= 4.0$~GeV. The other parameters were found by using the $\chi^2$ method in order to fit 
the experimental data  for $\cos\theta<0$ distribution from Ref.~\cite{expnew}.

\subsection{$KKK$ channel}\label{sec:kkk}

The CP asymmetry in the $B^-$ and $B^+$ event distributions in the $B^{\pm}\to K^{\pm} 
K^+K^-$ channel is compared to our model with the sum of the events of $\cos\theta<0$ 
and $\cos\theta>0$ as we did for $B^{\pm}\to \pi^{\pm} K^+K^-$ decay channel and the data 
provided by the LHCb in this last case. The functional form of the asymmetry for the 
$B^{\pm}\to K^{\pm} K^+K^-$ decay is the same as in Eq.~(\ref{dgpikk}). The only change 
is due to the kinematical factor $a'(s)$ in Eq.~(\ref{aprime}), that now presents the 
replacement $m_\pi\to m_K$.

\begin{figure}[!htb]
\centering
\psfrag{B --> KKK}[b][b][0.8]{$B^\pm\to K^\pm K^+K^-$}
{\includegraphics[scale=0.26]{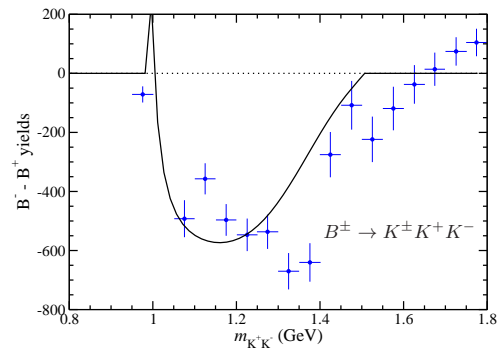}}
\caption{CP asymmetry of the \mbox{$B^\pm\to K^\pm K^+K^-$} decay compared with 
experimental values (blue points) taken from the sum of Figs.~6c and 6d of 
Ref.~\cite{expnew}.}
\label{b-kkk}
\end{figure}

We plot in Fig.~\ref{b-kkk} the asymmetry of the $B^{\pm}\to K^{\pm} K^+K^-$ decay, 
computed with the parameter $\mathcal{B}$ obtained in our previous fitting of the 
\mbox{$B^\pm\to K^\pm \pi^+\pi^-$} decay. Our model is compared to the data obtained by 
summing the experimental results of $B^{\pm}\to K^{\pm} K^+K^-$ distributions for both 
$\cos\theta$ positive and negative regions. 

\section{Concluding Remarks}\label{sec:finalremarks}

\subsection{Beyond the model}\label{sec:beyond}

We can go beyond the asymmetry formula in Eq.~(\ref{cp27d}), where we have not included 
the interference terms between the resonances and the scattering terms in the elastic 
channel, we remind that in the BW we included the elastic amplitude, but it is also 
important to consider it outside the resonance region, and returning to (\ref{cp27b}) as 
the starting point. We have neglected the Bose symmetrization of the amplitude by the 
exchange of the identical pions and the contribution of the double charm production and 
the its coupling to the $KK$ channel. This also should be addressed in the future.

The isoscalar and zero angular momentum $\pi\pi$ originated by contribution of the 
elastic scattering terms to the CP asymmetry for  \mbox{$B^\pm\to\pi^\pm\pi^+\pi^-$} and  
\mbox{$B^\pm\to K^\pm\pi^+\pi^-$}  decays written before, can be substituted the 
inelasticity and phase-shifts given by $\pi\pi\to KK$ S-matrix provided by 
Eq.~(\ref{s-matrix}) as $t^{J=0}_{\pi\pi,\pi\pi}= i \, \left( 1-\eta \, e^{2i 
\delta_{\pi\pi}} \right)$. The inclusion of the $S-$wave $\pi\pi$ scattering should 
improve the fitting below the $\rho$, where it is clearly seen in 
Fig.~\ref{b-pipipi-cos-neg}, the effect of a missing interference term. In this 
particular region the $f_0(600)$, namely the $\sigma$, will be taken into account by 
$t^{J=0}_{\pi\pi,\pi\pi}$, and we expect to improve the fit in the low mass region.

The interference between $\pi\pi \rightarrow KK$ and the $\rho$ resonance was used to 
fit the $K \pi\pi$ channel, as we can see in Fig.~\ref{b-kpipi-cos-neg}{\color{purple}}. 
Other possibility in this fit would be consider the large presence of $f_0(980)$, by 
using its interference with $\pi\pi \rightarrow KK$. This component do not satisfy CPT 
locally, but is consistent with Eq.~(\ref{cp27b}) with the coupled channel $B^\pm 
\rightarrow K^\pm K^+ K^-$. In this work, we choose to use only locally CPT invariant 
components for the interference with resonances.

\subsection{Conclusions}

We show how CP asymmetry in charmless three-body $B^\pm$ decays is constrained by CPT in 
the presence of resonances and  final state interaction in leading order. The CPT 
constraint $\sum_\lambda\Delta\Gamma_\lambda=0$, should include in the sum over the 
final states, all the kinematical allowed phase space  and possible decay channels. In 
leading order, the  scattering matrix $t_{\lambda,\lambda}$ corresponding to the decay 
channel where a resonance is formed, e.g., the $\rho$ in p-wave $\pi\pi$ elastic 
amplitude, accounts  for the dressing of the decay vertex $\rho\to\pi\pi$. The physical 
decay amplitudes associated with the partonic source, that give the $\rho$ meson and in 
its sequential decay to two pions, is dressed by the $\pi\pi$ re-scattering process in 
the resonant channel. The non-resonant channels receive contribution from the 
re-scattering in leading order. The decay of the resonance and re-scattering by the 
bachelor meson, should appear as a next-to-leading order contribution, to be consider in 
future works, where the general formulas for the three-body decay amplitude based on the 
CPT constraint that includes the formation of resonances, re-scattering and coupling to 
inelastic channels should be addressed. In our limited two-body description, we explored 
the rich patterns of the CP asymmetry formula, namely interference between non-resonant 
and resonant terms, as well as the coupling between the $\pi\pi$ and $KK$ channels due to 
the strong interaction.

We analysed in our study two examples of CP violation  in charmless $B^\pm$ three-body 
decays in the presence of resonances, excluding the possibility of channel coupling. 
The first case discussed is the $\rho(770)$ meson resonance plus a non resonant channel, 
and the second one is the case of two interfering  resonances, namely, $\rho(770)$ and 
$f_0(980)$. These examples discussed within two amplitude models are suitable for the 
study of $B^{\pm}\to \pi^{\pm} \pi^+\pi^-$  and  $B^{\pm}\to K^{\pm} \pi^+\pi^-$ for 
$\pi^\pm\pi^\mp$ masses below the $KK$ threshold, as provided by the new data LHCb 
data~\cite{expnew}, and we selected terms that are locally constrained by CPT.

The CP asymmetry produced by the $\rho(770)$ meson resonance in the simple two amplitude 
model with a non-resonant channel, which are brought by the real and imaginary parts of 
the interference term, presents a clear signature in the low mass $\pi^+\pi^+$ spectrum 
for both decays $B^{\pm}\to \pi^{\pm} \pi^+\pi^-$ and  $B^{\pm}\to K^{\pm} \pi^+\pi^-$, as 
indeed the new data suggests~\cite{expnew}. One term shows a very clear signature: a zero 
at the $\rho(770)$ mass in both regions of $\cos\theta<0$ and $\cos \theta <0$ for 
$B^{\pm}\to \pi^{\pm} \pi^+\pi^-$ with a change of the sign around $\cos \theta =0$.

The two amplitude model constituted by the resonances $\rho(770)$ and $f_0(980)$, shows 
two zeros in the CP asymmetry from the real part of the Breit-Wigners: one close to the 
$\rho(770)$ mass and another one close to the $f_0(980)$ mass. However, the term which 
mixes the real and imaginary parts of the Breit-Wigners does not present a clear signature 
of CP violation around the $\rho(770)$ meson mass, while a zero is present at the 
$f_0(980)$ mass, because the ratio $\Gamma_\rho\,m_\rho/\Gamma_f\,m_f$ is considerably 
larger than one. In both examples, the zero's of the asymmetry at the resonance mass are 
associated only with the Dalitz CP violation from the interference of two amplitudes and 
the real part of the Breit-Wigner, the contributions associated with the imaginary part
present a behavior with a maximum at the resonances masses. The recent experimental 
results~\cite{expnew} are strongly indicating that the CP asymmetry appearing in the low  
$\pi^+\pi^+$  mass region  comes from the real part of the Breit-Wigners in the 
interference between two amplitudes involving the $\rho$ meson resonance. 

In addition, the CP asymmetries in whole kinematical region below 
$m_{\pi^+\pi^-}\lesssim1.8$~GeV in $B^{\pm}\to \pi^{\pm} \pi^+\pi^-$ and $B^{\pm}\to 
K^{\pm} \pi^+\pi^-$ decays demand the coupling to the $ \pi^{\pm} \pi^+\pi^-$ by the 
strong interaction, as has already been discussed in Ref.~\cite{BedPRD14}. To account for 
that, we introduced in the fitting of the CPV, the contribution of the coupled $\pi\pi$ to 
$KK$ channels in isoscalar and zero angular states. This was relevant for the fitting  
above the $KK$ threshold. The independent confirmation of the CP asymmetry flowing between 
different channels coupled by the strong interaction was checked  by the fair reproduction 
of the CPV in $B^\pm\to\pi^\pm K^+K^-$ and $B^\pm\to K^\pm K^+K^-$ distributions. The 
computation of the  asymmetry was performed as dictated by the CPT constraint applied to 
coupled channels. There was no new parameters beyond those given in the fitting of the CP 
asymmetry data in the $B^\pm\to\pi^\pm \pi^+\pi^-$ and $B^{\pm}\to K^{\pm} \pi^+\pi^-$   
decays.

We discussed improvements to CP asymmetry formula beyond the one used to fit the 
experimental data on \mbox{$B^\pm\to\pi^\pm \pi^+\pi^-$} decay, where the interference  
between the resonances and elastic scattering terms are missing. The inclusion of the 
interference between the $S-$wave $\pi\pi$ scattering amplitude and the $\rho(770)$ meson 
Breit-Wigner should improve the fit below the $m_\rho$. In this particular region, the 
$f_0(600)$, namely the $\sigma$, will be taken into account through the elastic $\pi\pi$ 
scattering, which it is expected to improve the fit. Still the Bose symmetrization of the 
amplitude by the exchange of the identical pions has to be considered, which should be 
addressed in the future.

Finally, the results of our fittings suggest an important consequence to CP violation: 
the strong phases can come from the hadronic interaction minimizing the effect of the 
imaginary part of penguin contributions to both decays studied here, as the recent 
experimental results are indicating. 

\appendix

\section{Angle integrated asymmetry}
\label{app-int}

By defining in Fig.~\ref{b-pipipi-theta}, $\pi^+\equiv 1$, $\pi^-\equiv 2$, and 
${\pi^\prime}^+\equiv 3$ (bachelor), one has that $m_{\pi^+\pi^-}=m_{12}=\sqrt{s}$, and 
$m_{{\pi^\prime}^+\pi^-}=m_{32}$. The kinematics of the decay imposes that $\cos\theta$ is 
a function of $s$ and $m_{32}$, namely, 
\begin{eqnarray}
\cos\theta=a(s)m_{32}^2+b(s),
\label{costheta}
\end{eqnarray}
where,
\begin{eqnarray}
a(s) &=& 
\frac{1}{(s-4m_\pi^2)^{1/2}\left[\frac{(M_B^2-m_\pi^2-s)^2}{4s}-m_\pi^2\right]^{1/2}},
\end{eqnarray}
and
\begin{eqnarray}
b(s) &=& 
-\frac{M_B^2 + 3m_\pi^2 - s} 
{2(s-4m_\pi^2)^{1/2}\left[\frac{(M_B^2-m_\pi^2-s)^2}{4s}-m_\pi^2\right]^{1/2}},\quad
\end{eqnarray}
with $M_B=5.279$~GeV (the detailed derivation of this equation is relegated to the 
Appendix~\ref{app-cos}). Therefore, the asymmetry presented in Eq.~(\ref{dgamacos}) is 
actually a function of $s$ and $m_{32}$, due to $\cos\theta$. In order to eliminate 
$m_{32}^2$, we integrate $\Delta\Gamma_\lambda(s,m_{32}^2)$ in this variable, leading to 
the asymmetry $\Delta\Gamma(s)\equiv\int\Delta\Gamma_\lambda(s,m_{32}^2)dm_{32}^2$. We 
still separate the integration in two regions, namely, that of $\cos\theta<0$, leading to 
\begin{equation}
\Delta\Gamma(s)^{(\cos\theta<0)}=\int_{(m_{32}^2)_{-1}}^{-b/a}
\Delta\Gamma_\lambda(s,m_{32}^2)dm_{32}^2, 
\label{dg-cos-neg}
\end{equation}
and that of $\cos\theta>0$, leading to 
\begin{eqnarray}
\Delta\Gamma(s)^{(\cos\theta>0)}=\int_{-b/a}^{(m_{32}^2)_{+1}}
\Delta\Gamma_\lambda(s,m_{32}^2)dm_{32}^2.
\label{dg-cos-pos}
\end{eqnarray}
The quantities
\begin{eqnarray}
(m_{23}^2)_{-1}(s) = -\frac{1+b(s)}{a(s)},\quad (m_{23}^2)_{+1}(s) 
=\frac{1-b(s)}{a(s)}\quad
\end{eqnarray}
correspond, respectively, to $\cos\theta=-1$ and $\cos\theta=+1$, according to 
Eq.~(\ref{costheta}). The quantity $-b/a$ is related to $\cos\theta=0$.

\begin{widetext}
The integration in Eqs.~(\ref{dg-cos-neg}) and (\ref{dg-cos-pos}) generates the following 
asymmetry,
\begin{eqnarray}
&\Delta\Gamma(s)& = 
\frac{\mathcal{A}}{a(s)\sqrt{s-4m_\pi^2}\left(1+\frac{s}{\Lambda^2_{\lambda}}\right)^2}
+\frac{\mathcal{B}\cos[2\delta_{\pi\pi}(s)]\sqrt{1-\eta^2(s)}}{a(s)\sqrt{s-4m_\pi^2}
\left(1+\frac{s}{\Lambda^2_{\lambda}}\right)\left(1+\frac{s}{\Lambda^2_{\lambda^\prime}}
\right)} 
+ \frac{\mathcal{C}|F^{\rm BW}_\rho(s)|^2k^2(s)}{3a(s)}
\nonumber \\
&+& \frac{|F^{\rm BW}_\rho(s)|^2k(s)}{2\xi a(s)}
\left\{\frac{\mathcal{D}(m^2_\rho-s)}{1+\frac{s} {\Lambda^2_{\lambda}}} 
+\frac{\mathcal{D}^\prime\sqrt{1-\eta^2(s)}\left\{m_\rho\Gamma_\rho(s)\cos[2\delta_{\pi\pi
}(s)]-(m^2_\rho-s)\sin[2\delta_{\pi\pi}(s)]\right\}}{\sqrt{s-4m_\pi^2}
\left(1+\frac{s}{\Lambda^2_{\lambda^\prime}}\right)} \right.
\nonumber \\
&+& \left. \frac{\mathcal{E}m_\rho\Gamma_\rho(s)}{1+\frac{s} {\Lambda^2_{\lambda}}} 
+\frac{\mathcal{E}^\prime\sqrt{1-\eta^2(s)}\left\{(m^2_\rho-s)\cos[2\delta_{\pi\pi
}(s)]+m_\rho\Gamma_\rho(s)\sin[2\delta_{\pi\pi}(s)]\right\}}{\sqrt{s-4m_\pi^2}
\left(1+\frac{s}{\Lambda^2_{\lambda^\prime}}\right)}\right\} 
\nonumber \\
&+&\frac{|F^{\rm BW}_{\rho}(s)|^2|F^{\rm BW}_f(s)|^2k(s)}{2\xi a(s)}\times 
\nonumber\\
&\times&\left\{\mathcal{F}[(m_\rho^2 -s)(m_f^2-s) + m_\rho\Gamma_\rho(s)m_f\Gamma_f(s)] 
+\mathcal{G}[(m_\rho^2 -s)m_f\Gamma_f(s) - m_\rho\Gamma_\rho(s)(m_f^2-s)]\right\} 
\nonumber \\
&+& \frac{|F^{\rm BW}_f(s)|^2}{a(s)}
\left\{\frac{\mathcal{H}(m^2_f-s)}{1+\frac{s}{\Lambda^2_{\lambda}}} 
+\frac{\mathcal{H}^\prime\sqrt{1-\eta^2(s)}\left\{m_f\Gamma_f(s)\cos[2\delta_{\pi\pi
}(s)]-(m^2_f-s)\sin[2\delta_{\pi\pi}(s)]\right\}}{\sqrt{s-4m_\pi^2}
\left(1+\frac{s}{\Lambda^2_{\lambda^\prime}}\right)} \right.
\nonumber \\
&+& \left. \frac{\mathcal{P}m_f\Gamma_f(s)}{1+\frac{s} {\Lambda^2_{\lambda}}} 
+\frac{\mathcal{P}^\prime\sqrt{1-\eta^2(s)}\left\{(m^2_f-s)\cos[2\delta_{\pi\pi
}(s)]+m_f\Gamma_f(s)\sin[2\delta_{\pi\pi}(s)]\right\}}{\sqrt{s-4m_\pi^2}
\left(1+\frac{s}{\Lambda^2_{\lambda^\prime}}\right)}\right\} + \frac{\mathcal{Q}|F^{\rm 
BW}_f(s)|^2}{a(s)},
\label{dgama-final} 
\end{eqnarray}
\end{widetext}
where $\xi=-1$ for $\cos\theta<0$, and $\xi=1$ for $\cos\theta>0$. Here, the rest frame 
momentum of the pion, namely, $(s-4m_\pi^2)^{-1/2}$, is included in the terms presenting 
interference with the $\pi\pi \rightarrow KK$ amplitude, in order to introduce the 
kinematical factor in the scattering amplitude. Furthermore, this factor is consistent 
with the CPT constraint, which indicates that the decay widths for the coupled 
$\pi\pi\pi$ 
and $\pi KK$ channels above the $KK$ threshold is such that 
$\Delta\Gamma_{\pi\pi\pi}=-\Delta\Gamma_{\pi KK}$, as discussed at the end of 
Sec.~\ref{sec:pikk}. Finally, we also remind that all free parameters are compatible with 
the energy quantities of the model, that are in GeV units.

\section{Derivation of Eq.~(\ref{costheta})}
\label{app-cos}

In a general way, one has that
\begin{eqnarray}
P^\mu P^\prime_\mu &=& p^0p^\prime_0 + p^1p^\prime_1 + p^2p^\prime_2 + p^3p^\prime_3 
\nonumber \\
&=& EE^\prime - p_xp^\prime_x - p_yp^\prime_y - p_zp^\prime_z \nonumber \\
&=& EE^\prime - \vec{p}\cdot\vec{p}^{\,\prime} = p_0p^\prime_0 - 
\vec{p}\cdot\vec{p}^{\,\prime}.
\label{ap1}
\end{eqnarray}
If $P^\mu=P^\prime_\mu$, then Eq.~(\ref{ap1}) becomes
\begin{eqnarray}
P^\mu P_\mu = E^2 - \vec{p}^{\,2} = m^2 + \vec{p}^{\,2} - \vec{p}^{\,2} = m^2.
\end{eqnarray}

In the \mbox{$B^+\to \pi^+\pi^+\pi^-$} decay of Fig.~\ref{b-pipipi-theta}, we define 
$\pi^+\equiv 1$, $\pi^-\equiv 2$, and ${\pi^\prime}^+\equiv 3$ (bachelor). Therefore, 
$m_{\pi^+\pi^-}=m_{12}=\sqrt{s}$, $m_{{\pi^\prime}^+\pi^-}=m_{32}$, and 
$m_1=m_2=m_3=m_\pi$, since we have degenerated the pion mass. If we assume that
\begin{eqnarray}
P^\mu_{32}\equiv P^\mu_3 + P^\mu_2,
\end{eqnarray}
then, these definitions lead to
\begin{eqnarray}
m_{32}^2 &=& P^\mu_{32}\,P_{32\,\mu} = (P^\mu_3 + P^\mu_2)(P_{3\,\mu} + P_{2\,\mu}) 
\nonumber \\
&=& P^\mu_3P_{3\,\mu} + P^\mu_3P_{2\,\mu} + P^\mu_2P_{3\,\mu} + P^\mu_2P_{2\,\mu} 
\nonumber \\
&=& m_3^2 + p_3^0p_2^0 - \vec{p}_3\cdot\vec{p}_2 + p_2^0p_3^0 - \vec{p}_2\cdot\vec{p}_3 + 
m_2^2 \nonumber \\
&=& 2m_\pi^2 + 2p_3^0p_2^0 - 2\vec{p}_3\cdot\vec{p}_2 \nonumber \\
&=& 2m_\pi^2 + 2p_3^0p_2^0 + 2|\vec{p}_3||\vec{p}_2|\cos\theta.
\end{eqnarray}
Thus, $\cos\theta$ is written as
\begin{eqnarray}
\cos\theta = \frac{m_{32}^2 - 2m_\pi^2 - 2p_3^0p_2^0}{2|\vec{p}_3||\vec{p}_2|}.
\label{costheta-ap}
\end{eqnarray}

In order to write $\cos\theta$ in terms of $m_{32}^2$ and $m_{12}^2=s$, it is necessary 
to find $p_2^0$, $|\vec{p}_2|$, $p_3^0$, and $|\vec{p}_3|$. The first two quantities can 
be extracted as follows. Notice that,
\begin{eqnarray}
m_{12}^2 &=& s = P^\mu_{12}\,P_{12\,\mu} = (P^\mu_1 + P^\mu_2)(P_{1\,\mu} + P_{2\,\mu}) 
\nonumber \\
&=& P^\mu_1P_{1\,\mu} + P^\mu_1P_{2\,\mu} + P^\mu_2P_{1\,\mu} + P^\mu_2P_{2\,\mu} 
\nonumber \\
&=& m_1^2 + p_1^0p_2^0 - \vec{p}_1\cdot\vec{p}_2 + p_2^0p_1^0 - \vec{p}_2\cdot\vec{p}_1 + 
m_2^2 \nonumber \\
&=& 2m_\pi^2 + 2p_1^0p_2^0 - 2\vec{p}_1\cdot\vec{p}_2 \nonumber \\
&=& 2m_\pi^2 + 2p_1^0p_2^0 + 2|\vec{p}_1||\vec{p}_2|.
\end{eqnarray}
Furthermore, by the figure, we see that $|\vec{p}_1|=|\vec{p}_2|\equiv|\vec{p}_\pi|$. 
However, since $p_1^0=\sqrt{m_1^2+|\vec{p}_1|^2}$ and 
$p_2^0=\sqrt{m_2^2+|\vec{p}_1|^2}$, we conclude that $p_1^0=p_2^0\equiv 
p_\pi^0=\sqrt{m_\pi^2+|\vec{p}_\pi|^2}$. Then,
\begin{eqnarray}
s = 2m_\pi^2 + 2(m_\pi^2+|\vec{p}_\pi|^2) + 2|\vec{p}_\pi|^2 = 4m_\pi^2 + 
4|\vec{p}_\pi|^2,\qquad
\end{eqnarray}
with
\begin{eqnarray}
|\vec{p}_\pi| = \sqrt{\frac{s}{4}-m_\pi^2},
\label{ppi}
\end{eqnarray}
and
\begin{eqnarray}
p_\pi^0=\sqrt{m_\pi^2+|\vec{p}_\pi|^2}=\frac{\sqrt{s}}{2}.
\label{p0pi}
\end{eqnarray}

In order to find the $s$ dependence of $p_3^0$ and $|\vec{p}_3|$, we proceed to write
\begin{eqnarray}
P_B^\mu = P_1^\mu + P_2^\mu + P_3^\mu = P_{12}^\mu + P_3^\mu,
\end{eqnarray}
and make
\begin{eqnarray}
M_B^2 &=& P^\mu_B\,P_{B\,\mu} = (P^\mu_{12} + P^\mu_3)(P_{12\,\mu} + P_{3\,\mu}) 
\nonumber \\
&=& P^\mu_{12}P_{12\,\mu} + P^\mu_{12}P_{3\,\mu} + P^\mu_3P_{12\,\mu} + 
P^\mu_3P_{3\,\mu}\nonumber \\
&=& s + p_{12}^0p_3^0 - \vec{p}_{12}\cdot\vec{p}_3 + p_3^0p_{12}^0 - 
\vec{p}_3\cdot\vec{p}_{12} + m_3^2.\nonumber\\
\end{eqnarray}
In the reference frame in which $\vec{p}_{12}=\vec{p}_1+\vec{p}_2=0$, we have
\begin{eqnarray}
M_B^2 &=& s + 2p_{12}^0p_3^0 + m_3^2.
\end{eqnarray}
Furthermore, in this reference frame, $p_{12}^0=\sqrt{s+|\vec{p}_{12}|^2}=\sqrt{s}$, 
leading to 
\begin{eqnarray}
M_B^2 &=& s + 2\sqrt{s}p_3^0 + m_\pi^2.
\end{eqnarray}
From this equation we get,
\begin{eqnarray}
p_3^0 = \frac{M_B^2 - m_\pi^2 - s}{2\sqrt{s}}.
\label{p03}
\end{eqnarray}
Finally, from $p_3^0=\sqrt{m_3^2 + |\vec{p}_3|^2}$, we have
\begin{eqnarray}
|\vec{p}_3|&=&\sqrt{\left(p_3^0\right)^2 - m_3^2} \nonumber \\
&=& \sqrt{\frac{(M_B^2-m_\pi^2-s)^2}{4s}-m_\pi^2}.
\label{p3}
\end{eqnarray}

Inserting Eqs.~(\ref{ppi}), (\ref{p0pi}), (\ref{p03}) and (\ref{p3}) in 
Eq.~(\ref{costheta-ap}) we get,
\begin{eqnarray}
\cos\theta &=& \frac{m_{32}^2 - 2m_\pi^2 - \frac{M_B^2 - m_\pi^2 - s}{2}}
{2\left(\frac{s}{4}-m_\pi^2\right)^{1/2}
\left[\frac{(M_B^2-m_\pi^2-s)^2}{4s}-m_\pi^2\right]^{1/2}}\nonumber \\
&=& \frac{2m_{32}^2 - M_B^2 - 3m_\pi^2 + s}{2\left(s-4m_\pi^2\right)^{1/2}
\left[\frac{(M_B^2-m_\pi^2-s)^2}{4s}-m_\pi^2\right]^{1/2}}.\qquad
\label{costhetafinal-ap}
\end{eqnarray}

This same equation is obtained if we now treat the \mbox{$B^-\to \pi^-\pi^+\pi^-$} decay. 
In this case, the $B$ meson and the pions of Fig.~\ref{b-pipipi-theta} have its 
charges changed, and one defines $\pi^-\equiv 1$, $\pi^+\equiv 2$ and 
${\pi^\prime}^-\equiv 3$ (bachelor). Therefore, one has $m_{\pi^+\pi^-}=m_{21}=\sqrt{s}$, 
 and $m_{\pi^+{\pi^\prime}^-}=m_{23}$. For this case, $\cos\theta$ will be written 
exactly as in Eq.~(\ref{costhetafinal-ap}), but for $m_{32}\to m_{23}$.

\section*{Acknowledgments} 

We thank the support from Conselho Nacional de Desenvolvimento Cient\'ifico e 
Tecnol\'ogico (CNPq) of Brazil. O.~L. also acknowledges the support of the grant 
$\#$2013/26258-4 from S\~ao Paulo Research Foundation (FAPESP). J. H. A. N. acknowledges 
the support of the grant $\#$2014/19094-8 from S\~ao Paulo Research Foundation (FAPESP).

\end{document}